\documentclass[pra,aps,showpacs]{revtex4}
\usepackage{graphicx, amsmath,float}

\begin{document}

\title{Collective excitations in trapped boson-fermion mixtures: from
demixing to collapse}
\author{P. Capuzzi, A. Minguzzi, and M. P. Tosi}
\affiliation{NEST-INFM and Classe di Scienze, Scuola Normale Superiore,
I-56126, Pisa, Italy} 
\begin{abstract}
We calculate the spectrum of low-lying collective excitations in a
gaseous cloud formed by a Bose-Einstein condensate and a
spin-polarized Fermi gas over a range of the boson-fermion coupling
strength extending from strongly repulsive to strongly
attractive. Increasing boson-fermion repulsions drive the system
towards spatial separation of its components (``demixing''), whereas
boson-fermion attractions drive it towards implosion
(``collapse''). The dynamics of the system is treated in the
experimentally relevant collisionless regime by means of a
Random-Phase approximation and the behavior of a mesoscopic cloud
under isotropic harmonic confinement is contrasted with that of a
macroscopic mixture at given average particle densities. In the latter
case the locations of both the demixing and the collapse phase
transitions are sharply defined by the same stability condition, which
is determined by the softening of an eigenmode of either fermionic or
bosonic origin. In contrast, the transitions to either demixing or
collapse in a mesoscopic cloud at fixed confinement and particle
numbers are spread out over a range of boson-fermion coupling
strength, and some initial decrease of the frequencies of a set of
collective modes is followed by hardening as evidenced by blue shifts
of most eigenmodes. The spectral hardening can serve as a signal of
the impending transition and is most evident when the number of bosons
in the cloud is relatively large. We propose physical interpretations
for these dynamical behaviors with the help of suitably defined
partial compressibilities for the gaseous cloud under confinement.
\end{abstract}
\maketitle

%%%%%%%%%%%%%%
\section{Introduction}
%%%%%%%%%%%%%%

Dilute boson-fermion mixtures are currently being produced and studied
in several experiments by trapping and cooling vapors of mixed
alkali-atom isotopes \cite{Schreck2001a, Goldwin01a, Hadzibabic02a,
Roati02a,Modugno02a,Ferlaino02a}.  The large variety of combinations
of atomic species and of choices of magnetic sublevels, together with
the additional possibility of tuning the interactions by means of
Feshbach resonances provide a unique opportunity of investigating
extensively the effects of boson-fermion interactions, both in the
case of mutual repulsions and in the case of attractions.

The boson-fermion coupling strongly affects the equilibrium properties
of the mixture and can lead to quantum phase
transitions. Boson-fermion repulsions can induce spatial demixing of
the two components when the interaction energy overcomes the kinetic
and confinement energies \cite{Molmer98a}, and several configurations
with different topology are possible for a demixed cloud inside a
harmonic trap \cite{Akdeniz02a}. Although spatial demixing has not yet
been experimentally observed, the $^6$Li-$^7$Li experiments of Schreck
{\it et al.}  \cite{Schreck2001a} appear to be quite close to the
onset of the demixed state \cite{Akdeniz02b}. In the case of
attractive boson-fermion coupling the mixture becomes unstable against
collapse when the boson-induced fermion-fermion attractions overcome
the Pauli pressure \cite{Molmer98a,Miyakawa01a}. Collapse has been
experimentally observed in a $^{40}$K-$^{87}$Rb mixture
\cite{Modugno02a}. Fermion pairing into a superfluid state has also
been predicted in the case of attractive interactions
\cite{Efremov02a}, but for spin-polarized fermions this is expected to
happen in the $p$-wave channel at temperatures much lower than the
current experimental limit. This possibility will not be considered in
the present work.

The dynamical properties of boson-fermion mixtures have been
previously investigated theoretically mainly in the mixed phase, both
for the macroscopic homogeneous system \cite{Yip2001a, Search02a, Viverit02b} and in mesoscopic clouds inside a harmonic trap. In the
latter case these studies have used a sum-rule approach
\cite{Miyakawa00a,Liu03a}, perturbation theory \cite{Minguzzi00a}, or
a Random-Phase approximation (RPA) \cite{Capuzzi01b, Sogo02a,
Capuzzi02a}.

In this paper we investigate the effect of boson-fermion interactions
on the spectrum of collective excitations as the mixture approaches an
instability and look for the dynamical signatures of the approaching
instabilities.  In a previous paper \cite{Capuzzi03a} we have studied
the transition to demixing of a confined cloud in the collisional
regime and found an analytical condition for the dynamical transition
point. We focus here on the experimentally more relevant case of the
collisionless regime, where we adopt an RPA scheme. Within this
formalism it is possible to study at the same level the evolution of
the cloud towards demixing on one side and collapse on the other, thus
providing a unified picture of very different states. We characterize
the dynamical properties of the mixture not only by its spectrum but
also through its generalized partial compressibilities.

We comparatively examine a homogeneous mixture and a mixture in
external harmonic confinement. Whereas in the former case the RPA
spectra are symmetric under changing the sign of the boson-fermion
coupling, inside the trap we find different behaviors on going from the
case of boson-fermion repulsions to that of attractions.  This is due
to the fact that in the theory the transition in the macroscopic
homogeneous system is approached at {\it fixed density} of the two
species, while in the trapped mesoscopic cloud we work in the
realistic situation of {\it constant particle numbers} and {\it fixed
external confinement}.

The paper is organized as follows. In Sec.~\ref{sec2} we review the
conditions for demixing and collapse as obtained from static
considerations, and in Sec.~\ref{rpa} we introduce the RPA formalism
for the collective excitation spectrum. Sec.~\ref{sec_hom} describes
the predictions of the RPA for the case of a homogeneous system, while
Sec.~\ref{sec4} reports the results for the collective excitation
spectrum of a mixture under external harmonic confinement. Finally,
Sec.~\ref{sec5} contains a summary and our main conclusions.

%%%%%%%%%%%%%%
\section{Equilibrium state: conditions for demixing and collapse}
%%%%%%%%%%%%%%
\label{sec2}

We consider a dilute boson-fermion mixture at zero temperature.  The
Hamiltonian which describes the system is
\begin{eqnarray}
H&=&\sum_{\sigma=B,F} \int d^3r \,   \Psi^\dagger_\sigma \left (
-\frac{\hbar^2 \nabla^2}{2m_\sigma} + V_{ext}^\sigma({\mathbf r}) -\mu_\sigma \right)
\Psi_\sigma \nonumber \\ &+& \frac{g_{BB}}{2}\int d^3 r\,
\Psi^\dagger_B\Psi^\dagger_B\Psi_B\Psi_B +g_{BF}\int d^3 r\,
\Psi^\dagger_B\Psi^\dagger_F\Psi_F\Psi_B
\end{eqnarray}
where $\Psi_B$ and $\Psi_F$ are the usual bosonic and fermionic field
operators, $V_{ext}^{B,F}({\mathbf r})$ are the external confining
potentials, $\mu_{B,F}$ are the chemical potentials, and $m_{B,F}$ the
atomic masses. We have introduced the couplings $g_{BB}=4\pi\hbar^2
a_{BB}/m_B$ for boson-boson interactions and $g_{BF}=2\pi\hbar^2
a_{BF}/m_{r}$ for boson-fermion interactions, in terms of the $s$-wave
scattering lengths $a_{BB}$ and $a_{BF}$ and of the reduced
boson-fermion mass $m_r$. We will always assume $a_{BB}>0$. We are
here neglecting the fermion-fermion interactions since we are
considering a spin-polarized Fermi gas where $s$-wave collisions are
forbidden by the Pauli principle and higher-order collisional
couplings are negligible at the temperatures of present interest.  In
the dilute limit we also neglect the quantum depletion of the
condensate.

The equilibrium state of the mixture is described at mean field level
by the self-consistent solution of the Gross-Pitaevskii equation for
the condensate extended to include the boson-fermion interactions,
\begin{equation}
\left[-\frac{\hbar^2 \nabla^2}{2m_B} + V_{ext}^B({\mathbf r}) + g_{BB}\, n_B({\mathbf r}) +
g_{BF}\, n_F({\mathbf r})\right]  \Phi_B({\mathbf r})=\mu_B \Phi_B({\mathbf r})
\label{gpe}
\end{equation}
and by the Hartree-Fock equations for the fermions coupled to the
boson density,
\begin{equation}
\left [-\frac{\hbar^2 \nabla^2}{2m_F} + V_{ext}^F({\mathbf r}) + g_{BF\,} n_B({\mathbf r}) \right] \psi_i({\mathbf r}) =
\varepsilon_i  \psi_i({\mathbf r}).
\label{hf}
\end{equation}
In Eqs.~(\ref{gpe}) and (\ref{hf}) the condensate density is given by
\begin{equation}
n_B({\mathbf r})=|\Phi_B({\mathbf r})|^2
\end{equation}
and the fermion density by
\begin{equation}
n_F({\mathbf r})=\sum_i |\psi_i({\mathbf r})|^2\,\theta(\mu_F-\varepsilon_i).
\label{nf}
\end{equation}
The chemical potentials are obtained by imposing the normalization
conditions $N_{B,F}=\int d^3r\, n_{B,F}({\mathbf r})$ giving the
numbers of particles of the two species. Excited-state condensate wave
functions and Hartree-Fock orbitals will also be needed in the
construction of the RPA dynamical susceptibilities in Sec.\ \ref{rpa}.

\begin{figure}
\includegraphics[width=0.55\linewidth,clip=true]{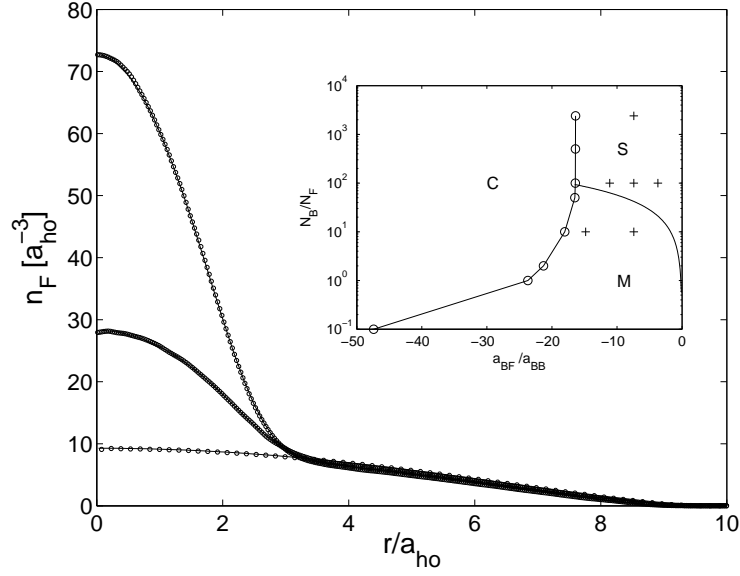}
\caption{\label{equilibrium} Equilibrium density profiles $n_F(r)$ (in
  units of $a_{ho}^{-3}$) as functions of the radial coordinate $r$
  (in units of $a_{ho}$) for the fermionic component of a
  $^6$Li-$^7$Li mixture under isotropic harmonic confinement, as
  obtained from the solution of the Hartree-Fock equations (dots) and
  from the Thomas-Fermi approximation (solid line, hidden by the
  dots). The curves from top to bottom correspond to the choices
  $a_{BF}=-4$ nm, $a_{BF}=-2$ nm, and $a_{BF}=0$. The other parameters
  are $N_B=10^5$, $N_F=10^4$, $a_{BB}=0.27$ nm and
  $\omega_0=2\pi\times 1000$ sec$^{-1}$. The inset reports the phase
  diagram of the mixture with boson-fermion attractions in the
  adimensional variables $a_{BF}/a_{BB}$ (linear scale) and $N_B/N_F$
  (log scale), showing the regions of stability (S), metastability (M)
  and collapse (C). The crosses show the points where the spectra of
  the trapped mixture are explored in this paper. }
\end{figure}

Under isotropic harmonic confinement
$V_{ext}^{B,F}(\mathbf{r})=m_{B,F}\omega_{B,F}^2 r^2/2$, for the
values of $N_F\sim 10^4$ as in current experiments we have verified
that the fermionic equilibrium profiles are well described by the
solution of the generalized Thomas-Fermi approximation (TFA)
\begin{equation}
-\frac{\hbar^2}{6m_F} \frac{\nabla^2 \sqrt{ n_F({\mathbf
r})}}{\sqrt{ n_F({\mathbf r})}}+  A n_F^{2/3}({\mathbf r})+
V_{ext}^F({\mathbf r}) + g_{BF\,} n_B({\mathbf r}) =\mu_F , 
\label{tfa}
\end{equation}
which includes the surface kinetic energy effects in the form of von
Weisz\"acker \cite{Capuzzi03a}. Here $A=(\hbar^2/2m_F)
(6\pi^2)^{2/3}$. The fermionic density profiles for various values of
the boson-fermion coupling $g_{BF}<0$ are shown in
Fig.~\ref{equilibrium}. The result of the TFA is practically
indistinguishable from the full solution of Eqs.\ (\ref{hf}) and
(\ref{nf}), since the shell effects play only a marginal role in the
spherically symmetric case for the above choice of fermion number. The
inset in Fig.\ \ref{equilibrium} reports the phase diagram of the
harmonically confined mixture for the case $g_{BF}<0$ (see below).

At increasing values of the boson-fermion coupling the mixture becomes
unstable against demixing (in the case $g_{BF}>0$) or collapse (in the
case $g_{BF}<0$) \cite{Molmer98a}.  In a homogeneous system at fixed
particle densities a linear stability analysis based on a mean-field
energy functional predicts the same condition for collapse and for
demixing \cite{Viverit00a}:
\begin{equation}
g_{BB} g_{FF}-g_{BF}^2 = 0
\label{hom_cond}
\end{equation}
with $g_{FF}=(2/3) A n_F^{-1/3}$ playing the role of an effective
fermion-fermion coupling due to the Pauli pressure in the Fermi gas.

In the presence of external harmonic confinement the effects of
 finiteness substantially modify the description of the
 instabilities. In the case of demixing the transition is smoothed out
 and the onset of demixing with increasing boson-fermion coupling is
 signalled first by a diminishing boson-fermion overlap energy
 \cite{Akdeniz02a}. The condition of such partial demixing is given by
\begin{equation}
\frac{g_{BF}^{part}}{g_{BB}} = \left(c_1\,\frac{N_F^{1/2}}{N_B^{2/5}} +
c_2\,\frac{N_B^{2/5}}{N_F^{1/3}}\right)^{-1},
\label{Ec:partdemix}
\end{equation}
where the parameters are $c_1=(15^{3/5}/48^{1/2})(m_F/m_B)^{3/2}
(a_{BB}/a_{ho})^{3/5}$ and $c_2 = (48^{1/3}/15^{3/5})(6/\pi)^{2/3}
(a_{BB}/a_{ho})^{2/5}$, with $a_{ho}=(\hbar/m_B\omega_B)^{1/2}$.  On
further increase of the boson-fermion coupling one reaches the point
where the fermionic density vanishes at the center of the trap, which
occurs at
\begin{equation}
\frac{g_{BF}^{dyn}}{g_{BB}}=\frac{\mu_F}{\mu_B}.
\label{Ec:sounddemix}
\end{equation}
This condition also corresponds to a sharp upturn in the frequencies
of collective excitations in the collisional regime \cite{Capuzzi03a}.
If the boson-fermion coupling is further increased the overlap between
the two clouds becomes negligible, the point of full spatial
separation being well predicted by Eq.\ (\ref{hom_cond}) in a
local-density approximation \cite{Akdeniz02b}. At full separation the
most likely configuration with current experimental parameters {\it
e.g.}~ in a $^6$Li-$^7$Li mixture \cite{Schreck2001a} is the ``egg''
configuration, formed by a core of bosons surrounded by a shell of
fermions.

In the case of attractive interactions, at increasing values of the
boson-fermion coupling the cloud is predicted to enter first a region
of metastability where the zero-point oscillations still prevent
collapse.  For equal masses and trapping frequencies the condition of
metastability as derived by Miyakawa {\it et al.} \cite{Miyakawa01a}
is given by
\begin{equation}
\frac{|g_{BF}^{meta}|}{g_{BB}}\ge \frac{N_B} {2^{5/2} \, N_F}.
\label{Ec:meta}
\end{equation}
On further increase of $|g_{BF}|$ the mixture reaches the point of
collapse, that in the numerical solution of the equilibrium equations
(\ref{gpe}) and (\ref{tfa}) we localize at the point where convergence
fails. We have verified that this instability point is well described
by Eq.~(\ref{hom_cond}) in a local-density approximation
\cite{Molmer98a,Miyakawa01a},
\begin{equation}
\frac{|g_{BF}^{coll}|}{g_{BB}}=\left[\frac{\pi m_B}{2 m_F}
\frac{1}{(6\pi^2 n_F(0))^{1/3} a_{BB}}\right]^{1/2}
\label{Ec:coll}
\end{equation}
where $n_F(0)$ is the fermion density at the center of the trap.  From
the conditions (\ref{Ec:meta}) and (\ref{Ec:coll}) it can be seen that
for large boson numbers $N_B \gg N_F$ the metastability region is
absent and the cloud is predicted to go directly from a stable to an
unstable configuration.  The inset in Fig.~\ref{equilibrium} shows the
phase diagram of the confined mixture in the case $g_{BF}<0$ with the
choice of parameters that we shall adopt later in the study of the RPA
spectrum of collective excitations. Similar curves for the critical
boson number needed to reach collapse have been reported by Roth
\cite{Roth02b}.

In summary, the criterion obtained in Eq.\ (\ref{hom_cond}) for a
macroscopic cloud is still useful to locate within a local-density
approximation the end of the transition to either demixing or collapse
in a boson-fermion mixture inside a harmonic trap. However, the
transitions in a mesoscopic cloud kept under fixed external
confinement are spread out and their approach will best be
characterized by dynamical signatures, as we proceed to show below by
spectral calculations in cases of increasing boson-fermion coupling.

%%%%%%%%%%%%%%
\section{The Random-Phase approximation for a boson-fermion
mixture}
\label{rpa}

The RPA is equivalent to a time-dependent Hartree-Fock theory of the
 dynamics of a quantum fluid in the collisionless regime
 \cite{Pines66a} and was first introduced for dilute boson-fermion
 mixtures under confinement by Capuzzi and Hern\'andez
 \cite{Capuzzi01b}. It treats on the same footing the density
 fluctuations of the two atomic species, satisfying the {\it f}-sum
 rules and allowing for Landau damping of the bosonic excitations by
 the Fermi cloud. Since it neglects correlations between density
 fluctuations, the RPA is valid in the dilute limit specified by the
 inequalities $n_B a_{BB}^3 \ll 1$ and $k_F |a_{BF}| \ll 1$, where
 $k_F=(6 \pi^2 n_F)^{1/3}$ is the Fermi wave number.

The RPA yields the spectrum of collective excitations in the linear
regime from a set of coupled equations for the density fluctuations
$\delta n_F $ and $\delta n_B$, which are obtained by assuming that
the fluid responds as an ideal gas to external perturbing fields
$\delta U_F$ and $\delta U_B$ {\it plus} the Hartree-Fock
fluctuations.  The RPA equations in Fourier transform with respect to
the time variable read
\begin{equation}
\delta n_F({\mathbf r},\omega)= \int d^3r' \, \chi^{0F}({\mathbf r},{\mathbf
r'},\omega) [\delta U_F({\mathbf r'},\omega)+ g_{BF}  \,\delta
n_B({\mathbf r'},\omega)]
\label{rpa_F}
\end{equation}
and
\begin{equation}
\delta n_B({\mathbf r},\omega)=\int d^3r' \, \chi^{0B}({\mathbf r},{\mathbf
r'},\omega) [\delta U_B({\mathbf r'},\omega)+ g_{BB}  \,\delta
n_B({\mathbf r'},\omega)+  g_{BF}  \,\delta
n_F({\mathbf r'},\omega)].
\label{rpa_B}
\end{equation}
In the inhomogeneous system consistency between equilibrium and
dynamics requires to take into account static mean-field interactions.
Thus $\chi^{0F}$ in Eq.\ (\ref{rpa_F}) is the Lindhard density-density
response function constructed with the Hartree-Fock orbitals from
Eq.~(\ref{hf}),
\begin{equation}
\chi^{0F}({\mathbf r},{\mathbf r'},\omega)=\sum_{i,j}
\frac{f(\varepsilon_i)-f(\varepsilon_j)}{\omega-(\varepsilon_j-\varepsilon_i)+i
\eta}
\psi_i({\mathbf r}) \psi^*_i({\mathbf r'})\psi_j ({\mathbf r'})\psi^*_j({\mathbf r})
\end{equation}
with $f(\varepsilon)=\theta(\mu_F-\varepsilon)$ and
$\eta=0^+$. Similarly, $\chi^{0B}$ in Eq.\ (\ref{rpa_B}) is the
condensate density-density response function constructed with the
excited states $\phi_i$ of the Gross-Pitaevskii equation with energy
eigenvalues $E_i$,
\begin{equation}
\chi^{0B}({\mathbf r},{\mathbf r'},\omega)= \sum_{i\neq 0} \left[\frac{\Phi_B ({\mathbf r})
\Phi_B^*({\mathbf r'})\phi_i^*({\mathbf r'}) \phi_i({\mathbf r})}{\omega-(E_i-\mu_B)+i\eta}-\frac{\Phi_B ({\mathbf r'})
\Phi_B^*({\mathbf r})\phi_i^*({\mathbf r}) \phi_i({\mathbf
r'})}{\omega+(E_i-\mu_B)+i\eta}\right].
\label{chi0b}
\end{equation}

The RPA Eqs.~(\ref{rpa_F}) and (\ref{rpa_B}) are most easily solved in
terms of a matrix of density-density response functions, which
are defined as
\begin{equation}
\chi_{\sigma\sigma'}({\mathbf r},{\mathbf r'},t-t')=\frac{\delta
n_\sigma ({\mathbf r},t)}{\delta U_{\sigma'} ({\mathbf r'},t')}= -i
\theta (t-t') \langle [\delta\rho_\sigma({\mathbf
r},t),\delta\rho_{\sigma'} ({\mathbf r'},t')] \rangle,
\end{equation}
where $\delta\rho_{\sigma}(\mathbf{r}, t)$ are the density
fluctuations operators in the Heisenberg representation. Upon 
introducing the matrices $\bar \chi =\left (
\begin{array}{cc} \chi_{FF} & \chi_{FB} \\ \chi_{BF} & \chi_{BB}
\end{array}\right)$, $\bar \chi^0 =\left ( \begin{array}{cc} \chi^{0F} 
& 0 \\ 0 & \chi^{0B} \end{array} \right)$, and $G =\left (
\begin{array}{cc} 0 & g_{BF} \\ g_{BF} & g_{BB} \end{array}\right)$
the RPA equations read
\begin{equation}
\bar \chi ({\mathbf r},{\mathbf r'},\omega)=\bar \chi^0 ({\mathbf
r},{\mathbf r'},\omega)+\int d^3r''\, \bar \chi^0 ({\mathbf
r},{\mathbf r''},\omega) \, G \,\bar \chi ({\mathbf r''},{\mathbf
r'},\omega) .
\end{equation}
 In fact, the boson-boson interactions can be resummed into the
 Bogoliubov response functions $\chi^{Bog}$, defined from
\begin{equation}
\chi^{Bog} ({\mathbf r},{\mathbf r'},\omega)= \chi^{0B} ({\mathbf
r},{\mathbf r'},\omega)+\int d^3r''\, \chi^{0B}({\mathbf r},{\mathbf
r''},\omega) \, g_{BB} \, \chi^{Bog} ({\mathbf r''},{\mathbf
r'},\omega) .
\end{equation}
Thus the RPA equation can alternatively be expressed as
\begin{equation}
\bar \chi ({\mathbf r},{\mathbf r'},\omega)=\tilde \chi^0 ({\mathbf
r},{\mathbf r'},\omega)+\int d^3r''\, \tilde \chi^0 ({\mathbf
r},{\mathbf r''},\omega) \, \tilde G \,\bar \chi ({\mathbf
r''},{\mathbf r'},\omega)
\label{Ec:rpabog}
\end{equation}
with $\tilde \chi^0 =\left ( \begin{array}{cc} \chi^{0F} & 0 \\ 0 &
\chi^{Bog} \end{array} \right)$ and $\tilde G =\left (
\begin{array}{cc} 0 & g_{BF} \\ g_{BF} & 0 \end{array}\right)$.  In
this way the excitations of the bosonic cloud are described as a gas
of Bogoliubov phonons interacting only with the fermions.

%%%%%%%%%%%%%%
\section{Spectra of the homogeneous mixture}
\label{sec_hom}

 The RPA equations for the homogeneous fluid mixture with average
particle densities $n_B$ and $n_F$ can be solved algebraically in
Fourier transform with respect to the relative position coordinate
$\mathbf{r}-\mathbf{r}'$. The solution takes the form
\begin{equation}
\bar \chi({\mathbf k},\omega)=\frac{1}{\epsilon({\mathbf k},\omega)}
\left ( \begin{array}{cc}
\chi^{0F}({\mathbf k},\omega)  & g_{BF}\, \chi^{Bog}({\mathbf k},\omega)\,\chi^{0F}({\mathbf k},\omega) \\ g_{BF}\, \chi^{Bog}({\mathbf k},\omega)\,\chi^{0F}({\mathbf k},\omega)
& \chi^{Bog}({\mathbf k},\omega) \end{array} \right)
\label{rpa_hom}
\end{equation}
where the ``dielectric function'' $\epsilon({\mathbf k},\omega)$ is
the common denominator of the four partial responses and is given by
\begin{equation}
\epsilon({\mathbf k},\omega)=1-g_{BF}^2 \,\chi^{Bog}({\mathbf k},\omega)\,\chi^{0F}({\mathbf k},\omega).
\label{epsilon}
\end{equation}
The spectrum of collective excitations is obtained by searching for
the poles of the response matrix or equivalently by imposing
$\epsilon({\mathbf k},\omega)=0$ in the complex frequency plane.  In
Eqs.~(\ref{rpa_hom}) and (\ref{epsilon}) the Bogoliubov response
function is
\begin{equation}
\chi^{Bog}({\mathbf k},\omega)=\frac{n_B^2
k^2/m_B}{\omega\,(\omega+ i \eta)-c_B^2k^2-(\hbar k^2/2m_B)^2} 
\label{chi_bog}
\end{equation}
with $c_B=(g_{BB}n_B/m_B)^{1/2}$, and the Lindhard response function is
\begin{equation}
\chi^{0F}({\mathbf k},\omega)= \sum_{\mathbf p} \frac{f(\varepsilon_{{\mathbf p}})-f(\varepsilon_{{\mathbf p}+{\mathbf k}})}{\omega-(\varepsilon_{{\mathbf p}+{\mathbf k}}-\varepsilon_{{\mathbf p}})+i\eta}
\label{rechif}
\end{equation}
where $\varepsilon_{\mathbf k}=\hbar k^2/2m_F$. The summation in Eq.\
(\ref{rechif}) can be carried out explicitly (see {\it e.g.} the book
of Fetter and Walecka \cite{Fetter71a}) to yield the spectrum of single
particle-hole pair excitations and the corresponding reversible part
of the susceptibility of the ideal Fermi gas.

Equations (\ref{rpa_hom}) and (\ref{epsilon}) were first introduced by
Yip \cite{Yip2001a}, who evaluated numerically the spectra in the case
of weak boson-fermion coupling. For $c_B>v_F$, where $v_F= \hbar
k_F/m_F$ is the Fermi velocity, the spectrum at $g_{BF}=0$ contains a
stable bosonic phonon and a fermionic particle-hole continuum. When
the boson-fermion interactions are turned on these modes hybridize and
repel each other, as can be analytically shown in the long wavelength
limit. From the condition $\epsilon({\mathbf k},\omega)=0$, using
Eq.~(\ref{chi_bog}) and $\chi^{0F}({\mathbf k},\omega)\rightarrow n_F
k^2/(m_F \omega^2)$ for $\omega > v_F k$ and $k \rightarrow 0$, we
obtain the correction to the Bogoliubov sound velocity due to
boson-fermion interactions as $c=c_B[1+g_{BF}^2 m_B n_F/ (g_{BB}^2 m_F
n_B)]^{1/2}$. This result agrees with that obtained by second-order
perturbation theory by Viverit and Giorgini \cite{Viverit02b}. In the
case $c_B<v_F$ the bosonic sound falls instead inside the fermionic
particle-hole continuum and its broadening by Landau damping must
therefore be considered.

\begin{figure}
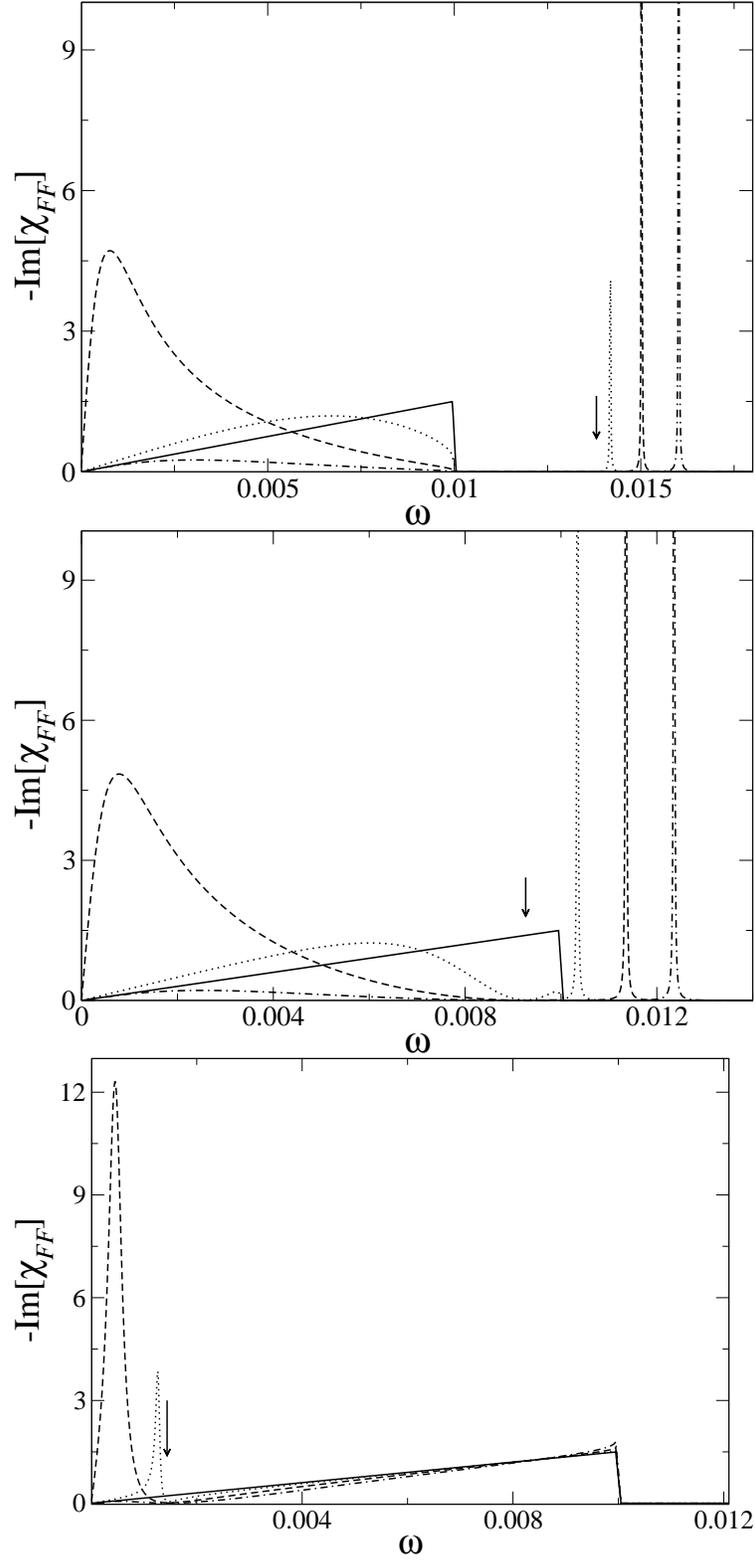

\includegraphics[width=0.55\linewidth,clip=true]{fig2_top.eps}
\includegraphics[width=0.55\linewidth,clip=true]{fig2_mid.eps}
\includegraphics[width=0.55\linewidth,clip=true]{fig2_bot.eps}
\caption{\label{fig1} Imaginary part of the fermion-fermion density
  response function $-$Im$\chi_{FF}({\mathbf k},\omega)$ for
  $k=0.01\,k_F$ as a function of $\omega$ (in units of $\hbar
  k_F^2/m_F$) in a homogeneous $^6$Li-$^7$Li mixture for various
  values for the coupling $a_{BF}$: the solid, dotted, dashed, and
  dash-dotted lines correspond to four equally spaced values of
  $|a_{BF}|$ from zero to the demixing or collapse point lying at
  $a_{BF}=7.9$ nm. The arrow marks the location of the Bogoliubov
  sound mode in the absence of boson-fermion interactions. The three
  panels from top to bottom correspond to the choices $c_B/v_F\simeq
  1.4$, $c_B/v_F\simeq 0.92$ and $c_B/v_F\simeq 0.14$. The
  corresponding values of the particle densities (in units of
  $a_{ho}^{-3}$) are $n_F=8.9$ and $n_B=32555$, 14649, and 351.5 from
  top to bottom. }
\end{figure}

At increasing values of the boson-fermion coupling one must resort to
a numerical solution of the RPA equations.  Figure \ref{fig1} shows
the fermionic spectrum ${\rm Im} \chi_{FF}({\mathbf k},\omega)$ for
various values of $g_{BF}$, the taking other parameters from the values at
the center of the trap in a $^6$Li-$^7$Li mixture \cite{Schreck2001a}
with equal trap frequencies $\omega_F=\omega_B=\omega_0=2\pi \times
1000\,$s$^{-1}$ and bosonic scattering length $a_{BB}= 0.27$ nm. While
in the RPA all four partial response functions
$\chi_{\sigma\sigma'}(\mathbf{k}, \omega)$ share the same poles, we
have chosen this spectrum because it shows most clearly the role of
the fermionic particle-hole continuum.

We consider first the case $c_B>v_F$. As the gas approaches the
instability points for demixing or collapse as given by Eq.\
(\ref{hom_cond}), we observe a softening of the particle-hole spectrum
until most of the oscillator strength is transferred to the bosonic
mode at the point of collapse or demixing (see the first panel in
Figure \ref{fig1}).  In the case $c_B\lesssim v_F$ we observe that the
boson-fermion interactions help to stabilize the bosonic mode, since
mode repulsion pushes it out of the particle-hole continuum (see the
second panel in Figure \ref{fig1}). This is not possible however in
the case $c_B\ll v_F$, where the effect of boson-fermion hybridization
is to break the particle-hole continuum into two parts (see the third
panel in Figure \ref{fig1}). In this case as the system approaches the
instability point it is the original bosonic mode which softens and
broadens.

The softening of the spectrum is directly reflected in the partial
``compressibilities'', obtained from the long wavelength limit of
the static susceptibilities as
\begin{equation}
 \chi_{\sigma\sigma'}(k\rightarrow 0,\omega=0)=\int_{-\infty}^{\infty}
\frac{d\omega'}{\pi} \frac{{\rm Im} \chi_{\sigma\sigma'}(k\rightarrow
0,\omega')}{\omega'} .
\end{equation}
In the homogeneous mixture these quantities can be calculated
analytically from the limiting behaviors of $\chi^{Bog}$ and
$\chi^{0F}$:
\begin{equation}
 \bar \chi(k\rightarrow 0,\omega=0)=
 \frac{1}{g_{BB}g_{FF}-g_{BF}^2}\left ( \begin{array}{cc} -g_{BB} &
   g_{BF} \\ g_{BF} & -g_{FF} \end{array} \right).
\label{compr}
\end{equation}
From the above equation we immediately see that the RPA predicts a
divergence in the compressibilities at the same instability points as
obtained from static considerations.  Alternatively, the instability
point of the homogeneous mixture can be predicted from the emergence
of unstable modes according to the condition
\begin{equation}
\lim_{\omega\rightarrow 0} \epsilon({\mathbf k}, i \omega)=0
\end{equation}
at long wavelengths.

Finally, from the form of the RPA solution in Eqs.\ (\ref{rpa_hom})
and (\ref{epsilon}) it is natural to define the effective
fermion-fermion and boson-boson coupling as $g_{eff}^{FF}=g_{BF}^2
\,\chi^{Bog}({\mathbf k},\omega)$ and $g_{eff}^{BB}=g_{BF}^2
\,\chi^{0F}({\mathbf k},\omega)$, respectively \cite{Bijlsma00a,
Heiselberg00a}. From this point of view the instabilities occur when
the effective interactions, which are negative in the limit
$(k\rightarrow 0,\omega = 0)$, overcome the Fermi pressure for the
fermions and the boson-boson repulsions for the bosons.

%In the next section we will estimate also the effective
%interactions for a trapped boson-fermion mixture.

%%%%%%%%%%%%%%
\section{Spectra of a mesoscopic cloud in a trap}
%%%%%%%%%%%%%%
\label{sec4}

We consider now a mixture under external harmonic confinement,
choosing for simplicity the spherically symmetric case
$V_{ext}^{B,F}=m_{B,F}\omega_0^2 r^2/2$ with the same trap frequency
$\omega_0$ for the two atomic species.  As for the homogeneous gas, we
investigate how the spectrum of collective excitations is modified as
the confined cloud moves toward demixing or collapse.  Important
differences arise from the presence of the confinement, under which
both the Bogoliubov sound and the fermionic particle-hole spectrum are
quantized. In particular the discrete energy levels depend on the
static Hartree-Fock fields and shift as we vary the coupling strength,
so that we know {\it a priori} that the spectra will not be symmetric
under a change in the sign of $g_{BF}$.

For the  finite system we follow the usual convention of
evaluating the spectral strength functions $\chi_{\sigma\sigma'}(\omega)$
\cite{Bertsch75a,Capuzzi01b}, which are defined as
\begin{equation}
\chi_{\sigma\sigma'}(\omega) =\int  d^3r  \int  d^3r' \,\delta
U_{\sigma'}^*(\mathbf{r})\,\chi_{\sigma\sigma'}(\mathbf{r},\mathbf{r'},\omega)
\delta U_{\sigma}(\mathbf{r'})
%\nonumber \\ &=&\int d^3r \, \delta
%U_{\sigma'}^*(\mathbf{r})\,\delta\rho_{\sigma}(\mathbf{r},\omega)
\end{equation}
where $\delta U_{\sigma}(\mathbf{r})$ are the intensities of the
time-dependent external potentials driving the two atomic species. In
the case of spherical confinement these have the form $\delta
U_{\sigma}(\mathbf{r})= f_{\sigma}(r) Y_{lm}(\hat{r})$ and
correspondingly we can decompose the response functions into
components of definite angular momentum,
\begin{equation}
\chi_{\sigma\sigma'}({\mathbf r},{\mathbf r}',\omega)=\sum_{l,m}
\frac{4\pi}{2l+1} \chi^{l}_{\sigma\sigma'}(r,r',\omega) Y_{lm}(\hat r)
Y_{lm}^*(\hat r'). 
\end{equation}
In these equations $Y_{lm}(\hat{r})$ are the spherical harmonic
functions and $\hat{r}$ denotes the unit vector along $\mathbf{r}$.

Phase separation into the ``egg'' configuration or collapse has
symmetry corresponding to $l=0$ and therefore we focus on the
monopolar excitations of the type $\delta U_{\sigma}(\mathbf{r}) =
U_0\,(r/a_{ho})^2 $. This type of perturbation can easily be produced in
current experiments since it corresponds to just a modulation of the
trap frequencies. In this case the strength function takes the form
\begin{equation}
 \chi_{\sigma\sigma'}^{l=0}(\omega)= U_0^2 \int d^3 r\int d^3r'\,
 \frac{r^2}{a_{ho}^2} \,\frac{{r'}^2}{a_{ho}^2}\,
 \chi_{\sigma\sigma'}^{l=0}(r,r',\omega) .
\end{equation}
We have obtained these RPA spectra by numerically solving the
integro-differential problem posed by Eqs.\
(\ref{rpa_F})-(\ref{chi0b}) in conjunction with the determination of
excited-state orbitals for the condensate and for the Fermi gas.  A
few points on the numerical solution are worth noting: ({\it i}) the
orbitals $\psi_i$ and $\phi_i$ are discretized on a uniform mesh of
about 250 points expanding to approximately twice the size of the
density profiles; ({\it ii}) convergence is achieved both in the
chemical potentials and in the respective energy spectrum; and ({\it
iii}) for attractive boson-fermion coupling the numerical solution is
computationally more demanding since a larger basis set is needed to
properly describe the equilibrium density profiles as we approach the
collapse.

\subsection{Across demixing}

We have calculated the RPA spectra of a mixture of $^7$Li and $^6$Li
atoms \cite{Schreck2001a} with trap frequency $\omega_0=2\pi \times
1000\,$s$^{-1}$, bosonic scattering length $a_{BB}= 0.27$ nm, and a
tunable $a_{BF}$ chosen to span the phase diagram from the mixed state
to partially demixed states (as we shall see below, the spectrum shows
a dynamical transition well before full separation).  Experimentally
the value of the boson-fermion scattering length can be tuned {\it
e.g.} by using a Feshbach resonance.

\begin{figure}
\includegraphics[width=0.5\linewidth]{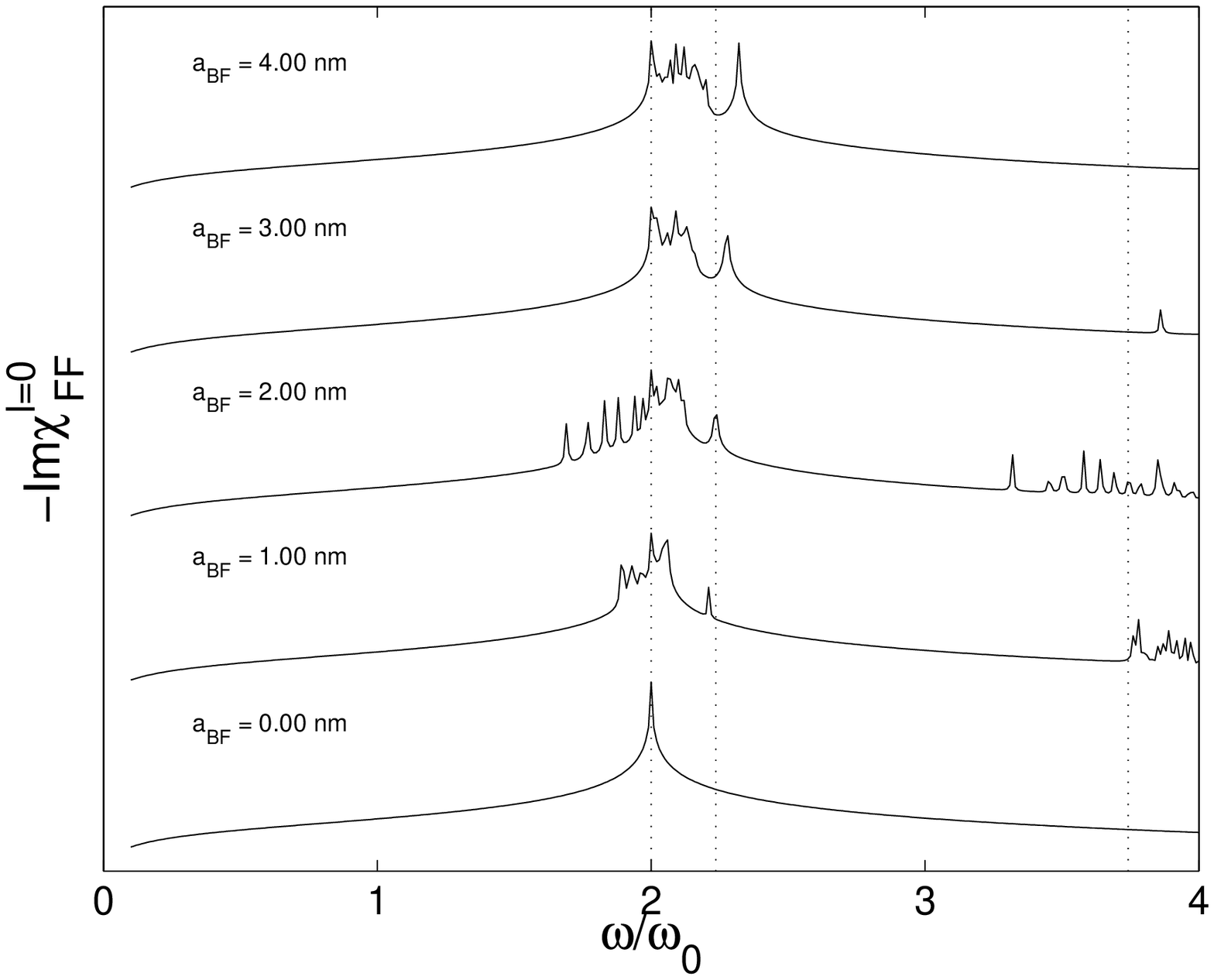}
\includegraphics[width=0.5\linewidth]{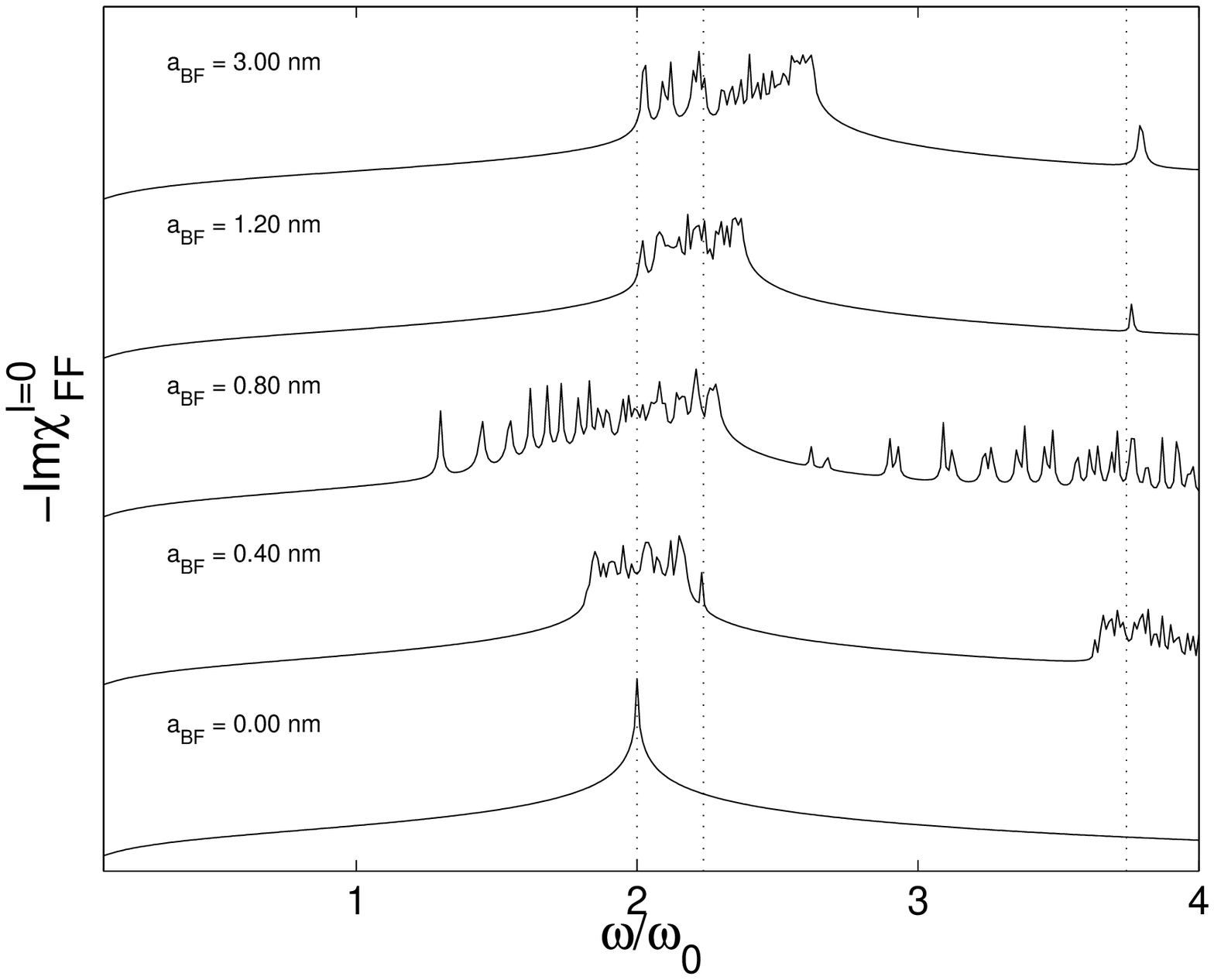}
\includegraphics[width=0.5\linewidth]{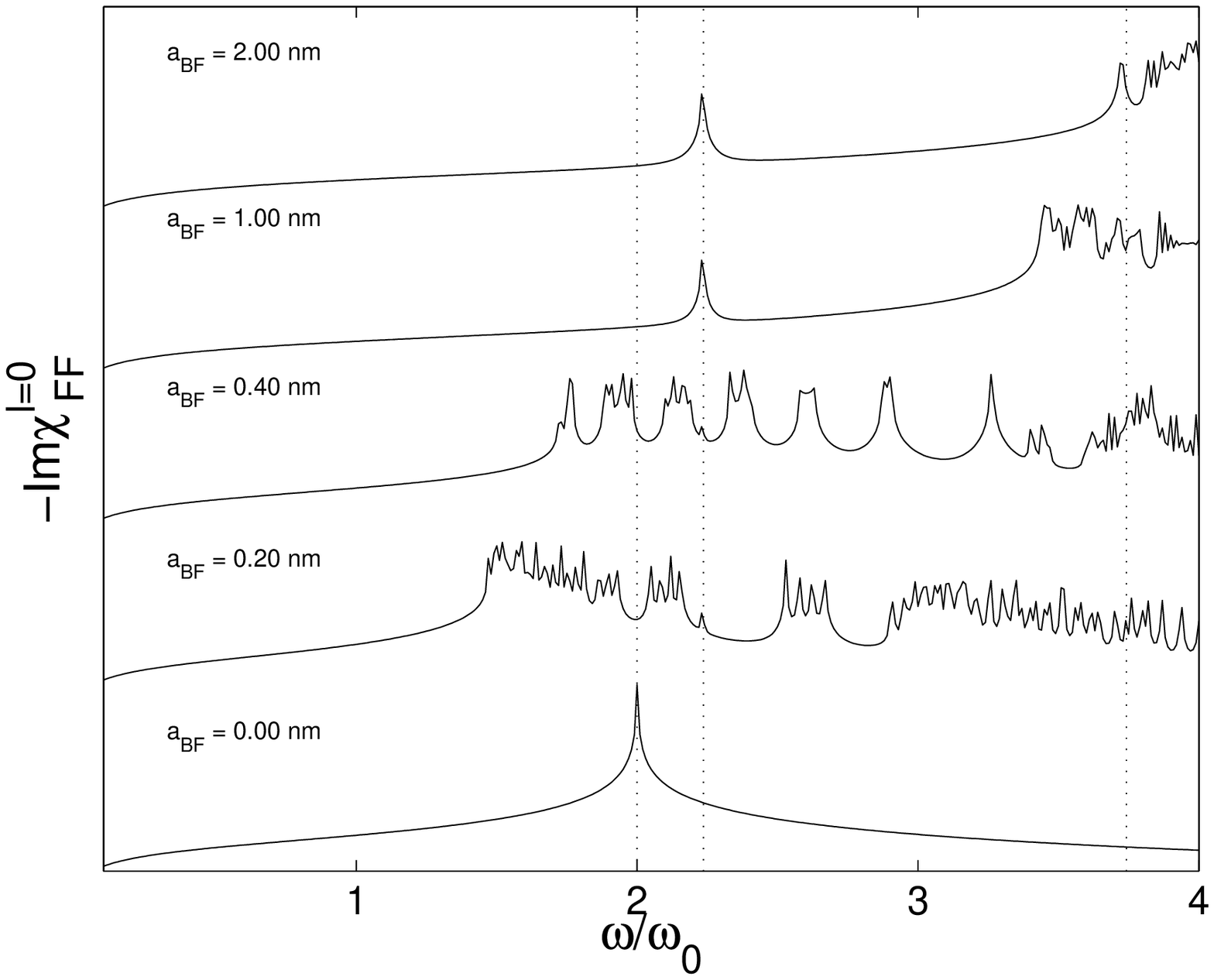}
\caption{\label{fig:spec0} Spectrum of monopolar response (in log
  scale and arbitrary units) as a function of frequency $\omega$ (in
  units of $\omega_0$) for a boson-fermion mixture in harmonic
  confinement with $N_F=10^4$ and $N_B = 10^5$ (top panel), $10^6$
  (middle panel), and $2.4\times10^7$ (bottom panel), for various
  values of $a_{BF}$ (in nm) as indicated in each panel.  The vertical
  dotted lines indicate the frequency of fermionic (left) and bosonic
  (right) modes in the absence of boson-fermion coupling.}
\end{figure}

In Fig.~\ref{fig:spec0} we show the imaginary part
Im${\chi}_{FF}^{l=0}(\omega)$ of the strength function for $N_F =
10^4$ and various values of $N_B$.  In the absence of boson-fermion
interactions the cloud responds to a monopolar excitation with only
two well-defined modes, a purely fermionic one at frequency
$\omega_f=2\,\omega_0$ \cite{Amoruso99a} and a purely bosonic one at
frequency $\omega_b \simeq \sqrt{5} \,\omega_0$
\cite{Stringari1996a}. The latter, of course, is not present in the
fermionic spectrum for $a_{BF}=0$.  Upon turning on the boson-fermion
interactions the bosonic peak appears in the fermionic strength
function with a considerable oscillator strength, showing that due to
the boson-fermion coupling the fermionic cloud can oscillate in
resonance with the bosons even in the collisionless regime.  This
``bosonic'' mode moves to higher frequency as $a_{BF}$ is increased,
analogously to what is found in the homogeneous mixture.  A second
``bosonic'' peak also appears in the fermionic strength function,
corresponding to a mode at $\omega_{b2}\simeq \sqrt{14}\, \omega_0$ in
the absence of boson-fermion coupling. This $n=2$ mode is not excited
by a monopolar perturbation in a pure bosonic cloud, but is excited in
the mixture due to the dynamical coupling between bosons and
fermions. Within the RPA the fermionic density fluctuations act as
further perturbing fields on the bosons, and {\it vice versa}.  .

For values of the frequency close to the original purely fermionic
peak we observe a strong fragmentation of the strength function, most
of which is already present at mean-field level in the Hartree-Fock
fermionic spectrum.  This fragmentation, which has also been reported
for other choices of system parameters \cite{Capuzzi01b,Sogo02a}, is
due to boson-fermion interactions lifting the angular degeneracy of
the single-particle levels for fermions in the trap. In addition to
fragmentation we observe a softening of the original fermionic
spectrum.  The softening stops when the topology of the cloud changes,
with the opening of a hole in the fermionic density at the trap
center, although strong overlap of the two clouds is still
present. This dynamical transition has also been found in the
collisional regime \cite{Capuzzi03a} and occurs in both regimes at the
same point, which is well described by the analytical expression
(\ref{Ec:sounddemix}). For larger values of the boson-fermion coupling
the original fermionic component of the spectrum moves upward, again
in analogy with the behavior of a trapped mixture in the collisional
regime \cite{Capuzzi03a}. This effect is intrinsically due to the
harmonic confinement: at increasing $a_{BF}$ the fermion cloud forms a
shell around the bosons with increasing average density and decreasing
thickness, thus increasing its stiffness in response to the monopolar
perturbation. In the homogeneous system instead the softening of the
fermionic spectrum continues up to the point of full thermodynamic
separation. The role of the number of bosons in the cloud is also
evident from Figure \ \ref{fig:spec0}.

As in Sec.~\ref{sec_hom}, it is possible to monitor the softening or
hardening of a part of the spectrum in the trapped cloud by looking at
the ``minus-one'' moment of the monopolar spectral strength
distribution. This spectral moment is specially sensitive to the
low-frequency part of the spectrum and, owing to the Kramers-Kronig
relations, yields the static limit
$\chi_{\sigma\sigma'}^{l=0}(\omega=0)$ of the strength function, thus
defining four generalized compressibilities which measure the reaction
of the cloud to a radial pressure exerted on a given component.  In
Fig.~\ref{fig:cmp_static} (left panels) we show the fermionic
compressibility $\chi_{FF}^{l=0}(\omega=0)$ both for positive and
negative values of $g_{BF}$ and for three choices of the number of
bosons. For positive $g_{BF}$ we observe an initial increase of
$\chi_{FF}^{l=0}(\omega=0)$, followed by a slow decrease after the
dynamical transition has taken place. The position of the maximum
corresponds to the dynamical point of demixing in
Eq.~(\ref{Ec:sounddemix}).

At variance from the homogeneous case, the four compressibilities in
the trapped cloud have a nonsymmetric behavior for positive and
negative values of the boson-fermion coupling. In particular this
means that the compressibilities cannot be predicted by a simple
local-density approach. For example, in the homogeneous gas one always
finds a negative value of -$\bar{\chi}_{BF}(\omega=0)$ for $g_{BF}>0$
irrespectively of the actual values of the densities, whereas as is
shown in Fig.\ \ref{fig:cmp_static} (right panels) this
compressibility becomes positive in the confined cloud after the
dynamical demixing point.

\subsection{Towards collapse}

We consider now the case of attractive boson-fermion interactions,
choosing the trap parameters of the $^6$Li-$^7$Li mixture as in the
previous section.  As is shown in the inset in Fig.~\ref{equilibrium},
when the numbers of bosons and fermions are comparable the cloud can
be found in a metastable state, while for larger numbers of bosons a
wide stable region is present in the phase diagram. The collapse is
reached at smaller values of $a_{BF}$ as the number of bosons is
increased, as can be understood from the fact that the boson-mediated
fermion-fermion attraction increases with $N_B$.

In the homogeneous gas the RPA spectra for attractive boson-fermion
interactions coincide with those given in Fig.~\ref{fig1} for the case
of repulsive interactions. This does not apply to the trapped mixture,
as can already be seen from the compressibilities in
Fig.~\ref{fig:cmp_static}. Contrary to the behavior of the homogeneous
gas where the generalized compressibilities diverge at the instability
points (see Eq.~(\ref{compr})), in the harmonic trap for negative
values of $a_{BF}$ they decrease and for large numbers of bosons tend
to vanish as the cloud approaches collapse. This is an indication that
spectral weight is being transferred to high-frequency modes, and the
effect becomes more dramatic as the number of bosons in the mixture is
increased. In the limit $N_B\gg N_F$ the decrease of the
fermion-fermion compressibility can be analytically predicted using a
single-mode approximation (see the inset in
Fig.~\ref{fig:cmp_static}). The details are given in Appendix
\ref{appA}.

A blue shift of the modes in the fermionic spectrum is found already
 at the Hartree-Fock level, as is illustrated in Fig.~\ref{fig:hist}
 by reporting the density $\rho_2(\varepsilon)$ of excited pairs,
\begin{equation} 
\rho_2(\varepsilon)=\sum_{n,l}[f(\varepsilon_{n,l})-f(\varepsilon_{n+1,l})]\,
\delta[\varepsilon-(\varepsilon_{n+1,l}- \varepsilon_{n,l})]
\label{rho2}
\end{equation}
 at increasing $N_B$ for $N_F=10^4$ and various values of $a_{BF}$. In
 Eq.~(\ref{rho2}) we have chosen only the pairs with $\Delta l=0$ to
 obtain a monopolar particle-hole excitation spectrum, with $\Delta
 n=1$ since this gives the most important contribution at low
 frequencies.

The spectra resulting from a full RPA calculation of the monopolar
fermion-fermion response in the region of boson-fermion attractions
are shown in Figs.~\ref{fig:spec2} and~\ref{fig:spec1}.  When the
numbers of bosons and fermions are comparable we observe a shift of
the pure fermionic peak towards lower frequencies. We interpret this
shift as being associated with the lifting of the degeneracy of the
single-particle Hartree-Fock spectrum by boson-fermion attractions,
which allow transitions between states that are closer in energy.  At
increasing values of $N_B$ we find instead that the fermionic mode is
blue-shifted by a much larger amount, as a result of the static
Hartree field of the bosons. The amount of the shift can be predicted
analytically in the limit $N_B\gg N_F$, where the interactions with
the bosons simply renormalize the trap frequency (see Appendix
\ref{appA}).  In the same limit the bosonic modes are almost
unaffected by the interactions with the fermions, as the fermionic
Hartree field becomes almost negligible for the bosons.

%%%%%%%%%%%%%%
\section{Summary and concluding remarks}
%%%%%%%%%%%%%%
\label{sec5}

In summary, in this paper we have studied how the mutual interactions
of bosons and fermions affect the spectrum of collective excitations
of a dilute boson-fermion mixture in the collisionless regime as the
mixture goes towards either demixing or collapse. To this purpose we
have used the Random-Phase approximation, which is most suited in the
dilute regime since it accounts correctly for the limit of small
couplings and generally conserves the $f$-sum rule.  We have studied both
a homogeneous mixture and a mixture in isotropic harmonic confinement,
where new effects arise due to the static mean-field interactions.  We
have indeed found that both the boson-fermion coupling and the
confinement strongly change the spectra of collective excitations. Our
main results are as follows.

First of all, in the presence of boson-fermion interactions the
excitation of one component of the mixture will set into motion also
the other component. There also is hybridization of the excitation
modes, {\it e.g.} the fermionic component can also oscillate in
resonance with the bosonic cloud at the frequency of a bosonic mode
and {\it vice versa}.

In the homogeneous gas we observe qualitatively different behaviors
depending on the ratio between the Bogoliubov sound velocity $c_B$ and
the Fermi velocity $v_F$. In the case $c_B>v_F$ at increasing values
of the boson-fermion coupling we find mode repulsion between the broad
particle-hole continuum and the Bogoliubov sound, and the fermionic
spectrum softens as the gas approaches either instability point. When
$c_B$ is just below $v_F$ the Bogoliubov sound is damped at low
couplings, but a ``revival'' of this mode can take place as the
boson-fermion interactions are increased. Softening and damping of the
bosonic mode is instead observed when $c_B \ll v_F$. Within the RPA we
have also derived a dynamical condition for the two instabilities,
which coincides with that obtained by studying the linear stability of
the mean-field energy functional.

In the trapped system we have studied the $l=0$ monopolar
excitations. While in the absence of interactions the spectrum of
density fluctuations shows only two sharp and independent modes, the
boson-fermion coupling induces fragmentation of the original fermionic
mode, revealing the presence of many occupied single-particle levels
in the Fermi sphere. At increased repulsive couplings we observe a
softening of the spectrum up to the point where the topology of the
fermionic cloud changes by the formation of a hole at its center. The
spectrum is then shifted to higher values of the frequencies at larger
values of $a_{BF}$.  For negative values of the coupling we observe
some frequency decrease for comparable numbers of bosons and fermions,
but as the number of bosons increases the static Hartree field of the
bosons drives a sizable blue shift of the fermionic modes.

From the comparative analysis of the collective excitation spectra of
the mixture in the homogeneous state and inside a trap we have learnt
that the inhomogeneous trap potential can change quite drastically the
physical picture, since the instabilities are approached along
thermodynamically different routes. In the homogeneous gas one works
at fixed average particle densities, while the confined cloud must be
treated at fixed external confinement. We have found that the partial
compressibilities cannot be described by a local-density picture, and
while they are expected to diverge in the homogeneous gas on
approaching the instability points, in the trap they saturate to
finite values as the component of the mixture are spatially
separating and vanish as collapse is approached. This suggests that a
description of boson-induced fermion-fermion interactions, {\it e.g.}
in the study of the instability against pairing, may also be affected
by the confinement.

In this work we have neglected the static and dynamical effects of
correlations and the quantum depletion of the condensate
\cite{Albus02a,Viverit02b}. These should become important in the
proximity of collapse as the gas densities increase. The effects of
temperature may also be of interest, with the thermal depletion of the
condensate and the blurring of the Fermi surface yielding, for
instance, damping of the Bogoliubov sound. Finally, new dynamical
effects may arise when the components of the trapped mixture have very
different atomic masses and will be studied elsewhere.

\acknowledgments
We acknowledge support from INFM  through the PRA-Photonmatter Program.

\appendix

\section{Single-pole model for the trapped mixture}
\label{appA}

We derive in this Appendix a single-pole model to describe the partial
fermion-fermion compressibility for attractive boson-fermion
interactions in the limit of large boson numbers. This model accounts
rather well for our numerical results and shows how the inhomogeneous
harmonic confinement in combination with the boson-fermion coupling
can give rise to a completely different behavior in the
compressibility relative to the homogeneous case.

In the limit $N_B \gg N_F$ the dynamical coupling between bosons and
fermions is quite small, but the fermionic density is largely affected
by the static mean field of the bosons. The bosonic equilibrium
profile is only weakly affected by the presence of the fermions, so
that in the Thomas Fermi approximation we can write
$n_B(\mathbf{r})=[\mu_B-V_{ext}^B(r)]\theta (R_B-r)/g_{BB}$ with
$R_B=(2\mu_B/m_B \omega_0^2)^{1/2}$.  In these conditions it is useful
to introduce an effective potential acting on the fermions as
\cite{Vichi00a}
\begin{eqnarray}
V_{eff}^{F}(r) &=&\frac{1}{2}m_F \omega_0^2 r^2 + g_{BF}
n_B(r)\nonumber \\
&\simeq & \frac{1}{2} m_F \omega_0^2 r^2 \left[1-\frac{m_B
g_{BF}}{m_F g_{BB}}\theta (R_B-r)\right]+ \frac{g_{BF}}{g_{BB}}\mu_B
\theta (R_B-r) .
\end{eqnarray}
For the choice $N_B=2.4\times 10^7$ and $N_F=10^4$ the bosonic and
fermionic clouds have the same radii, so that the effective potential
is seen by the fermions at all $r$ and acts as a harmonic trap with
renormalized frequency $\tilde \omega = \omega_0 [1-(m_B g_{BF})/(m_F
g_{BB})]^{1/2}$.

\begin{figure}
\centering
\includegraphics[width=0.45\linewidth,clip=true]{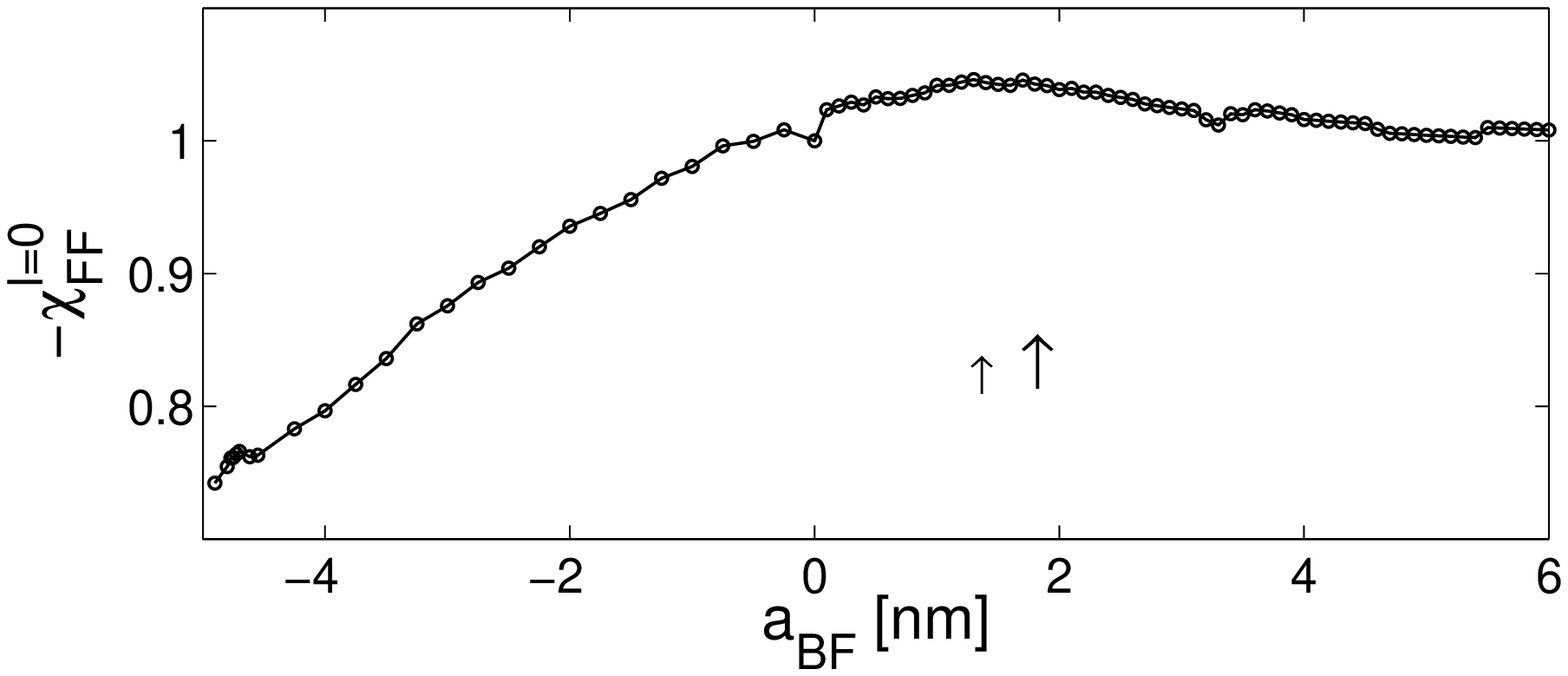}
\includegraphics[width=0.45\linewidth,clip=true]{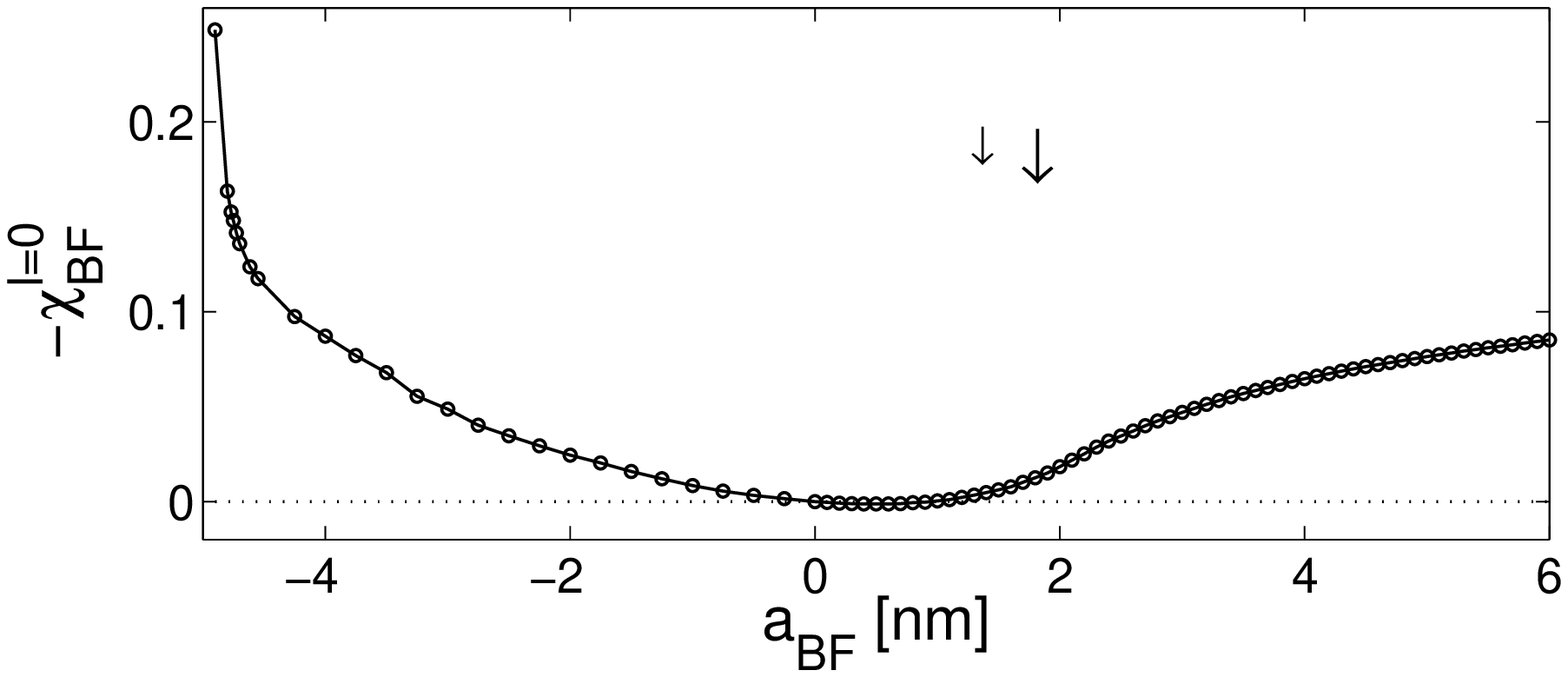}\\
\includegraphics[width=0.45\linewidth,clip=true]{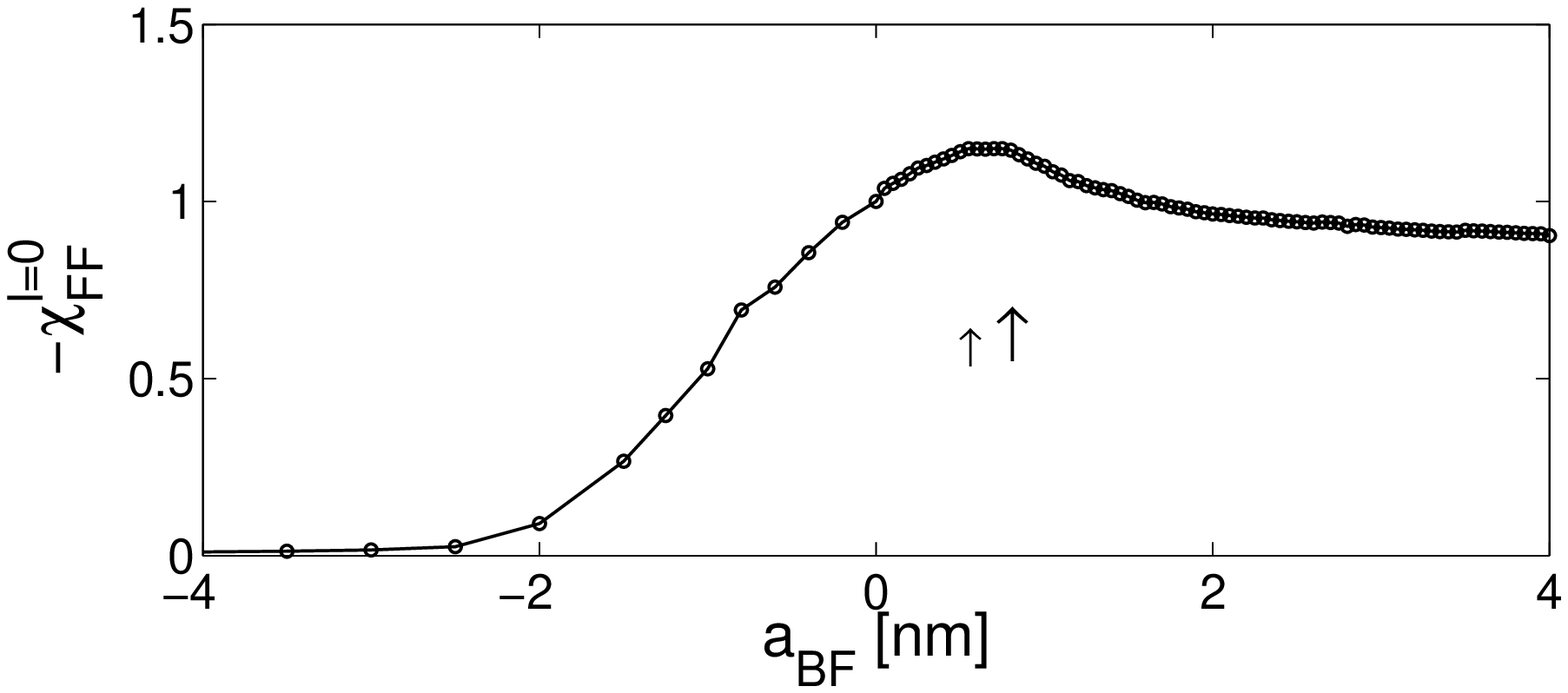}
\includegraphics[width=0.45\linewidth,clip=true]{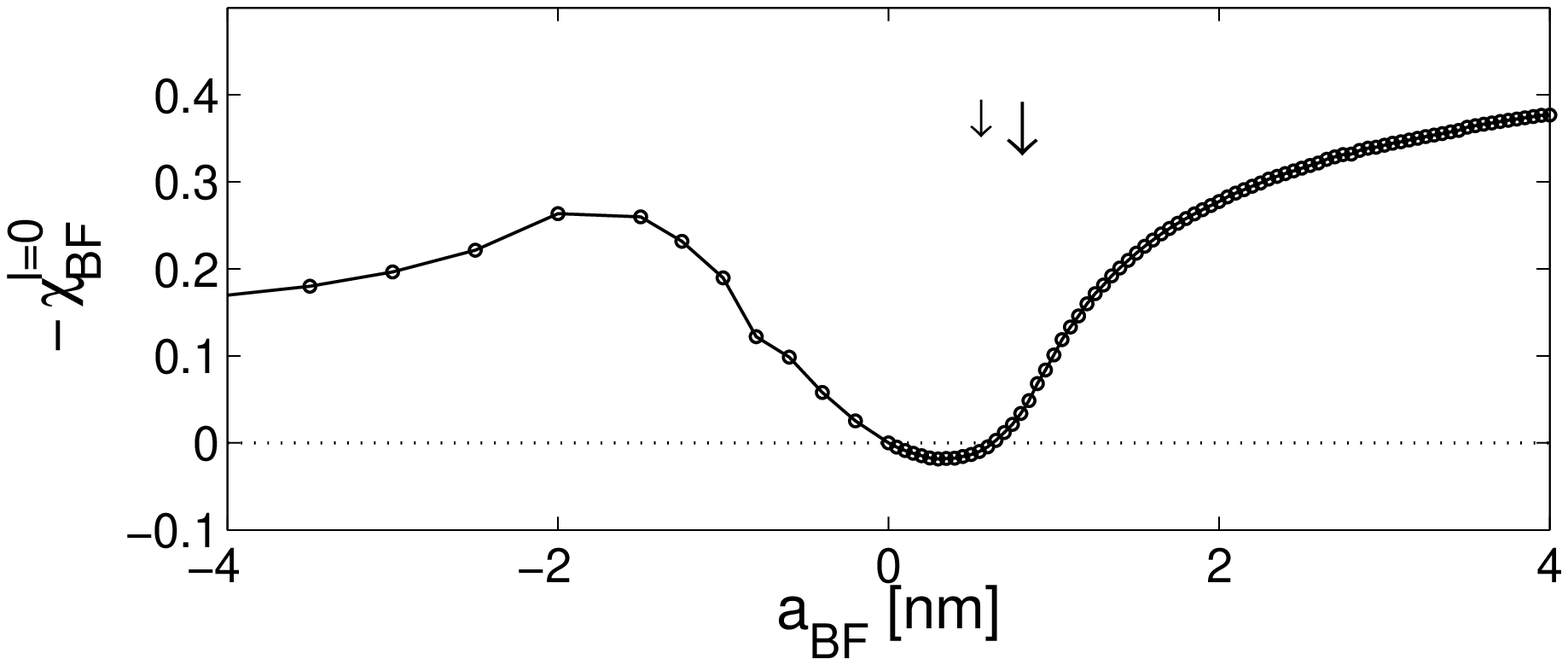} \\
\includegraphics[width=0.45\linewidth,clip=true]{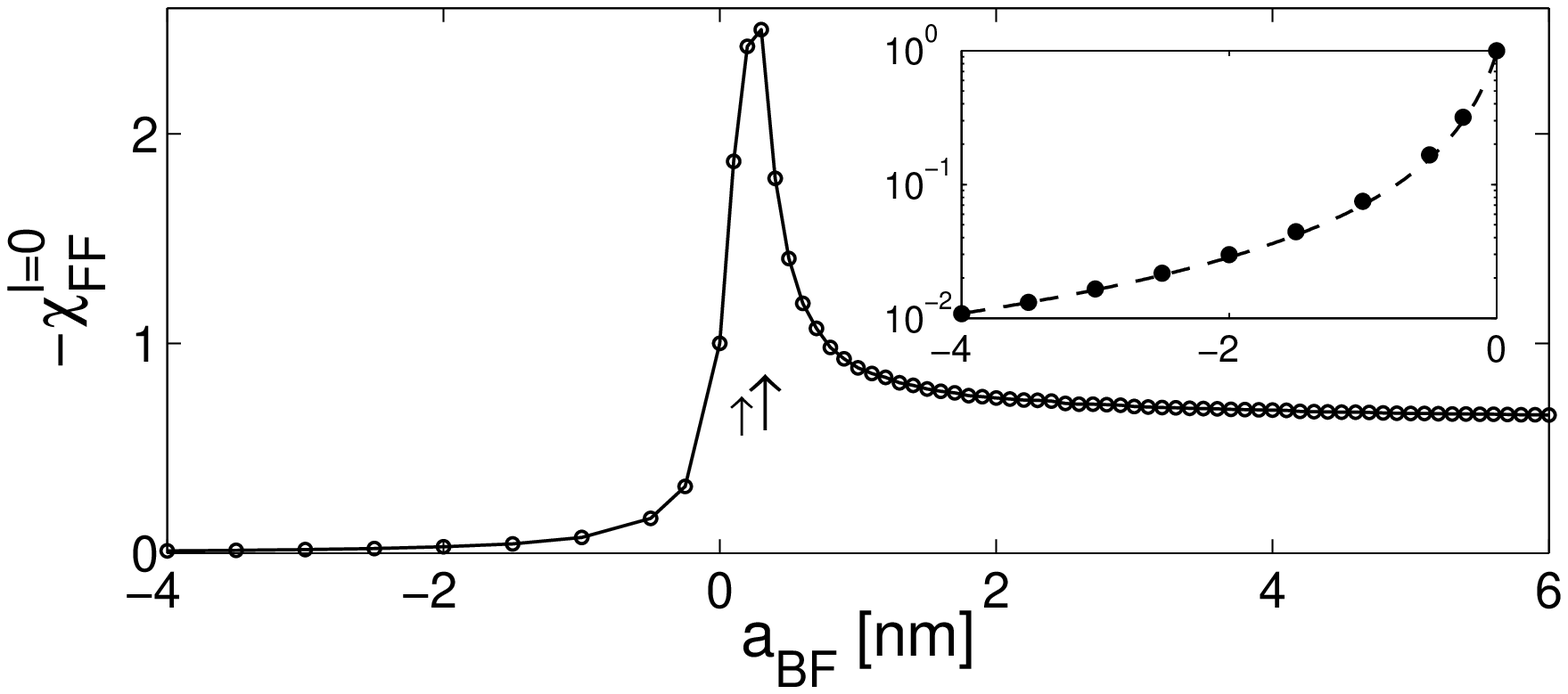}
\includegraphics[width=0.45\linewidth,clip=true]{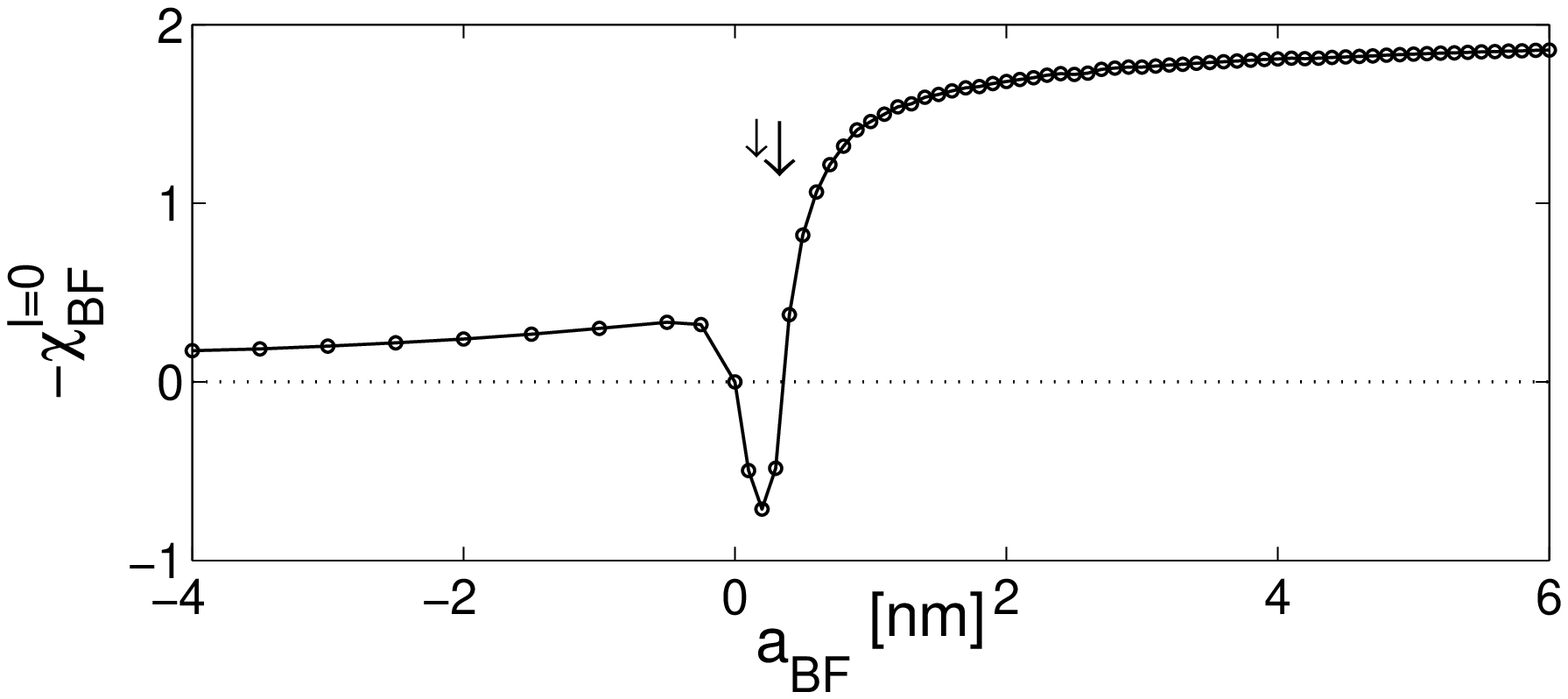}
\caption{\label{fig:cmp_static} Generalized compressibilities (in
  units of $\chi_{FF}^{l=0}$ at $a_{BF}=0$) of a boson-fermion mixture in harmonic confinement
  as functions of the boson-fermion scattering length $a_{BF}$ (in nm)
  for $N_F=10^4$ and $N_B = 10^5$ (top), $10^6$ (middle), and
  $2.4\times10^7$ (bottom). The small and large arrows indicate the
  points of onset of partial demixing and of the dynamical transition
  from Eqs.~(\ref{Ec:partdemix}) and (\ref{Ec:sounddemix}),
  respectively. The lines are a guide to the eye. The inset shows an
  enlargement of the region with $a_{BF}<0$ in comparison with the
  analytical predictions from Eq.~(\ref{a5}) (dashed line).}
\end{figure}

\begin{figure}
\includegraphics[width=0.5\linewidth]{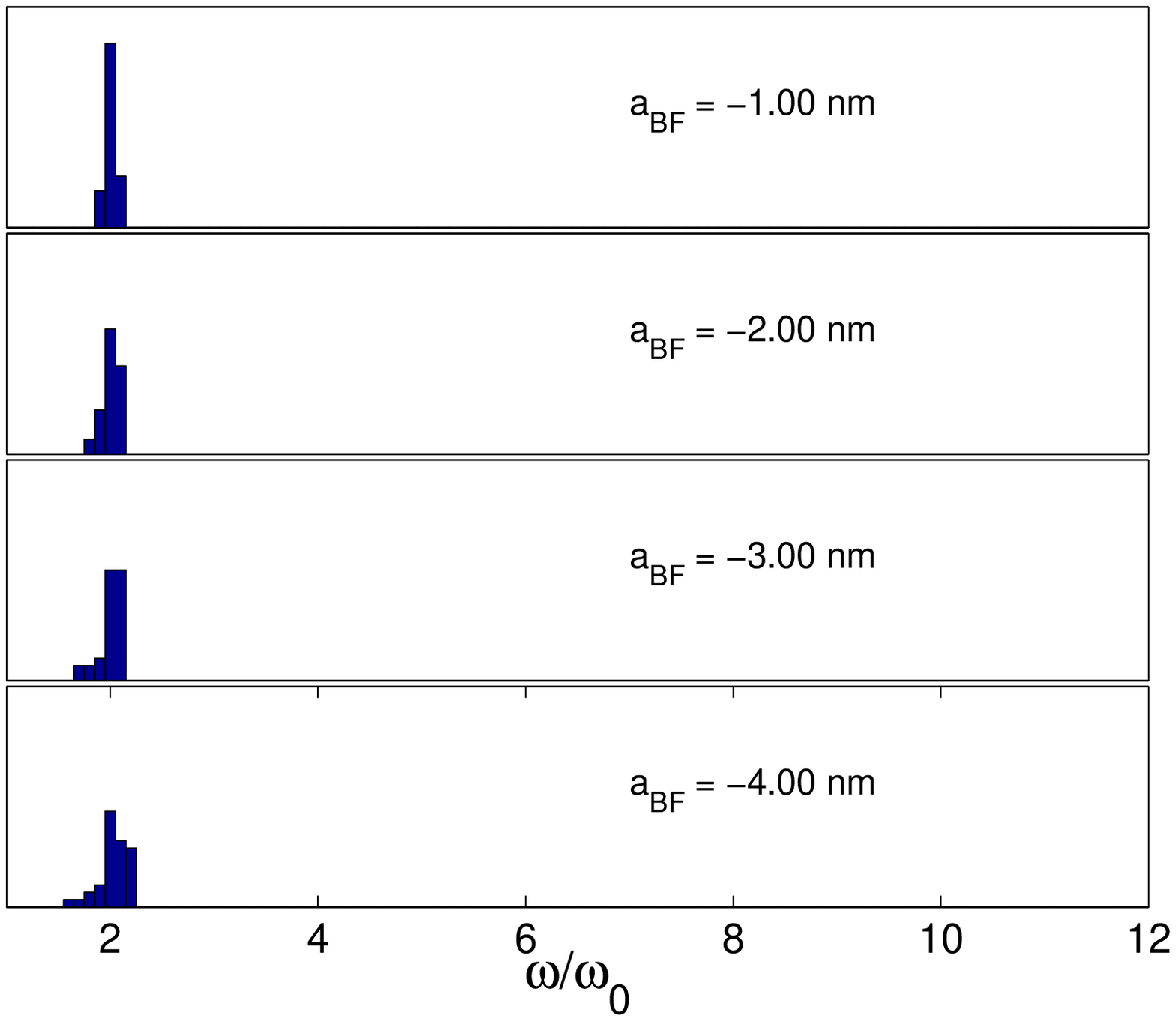}
\includegraphics[width=0.5\linewidth]{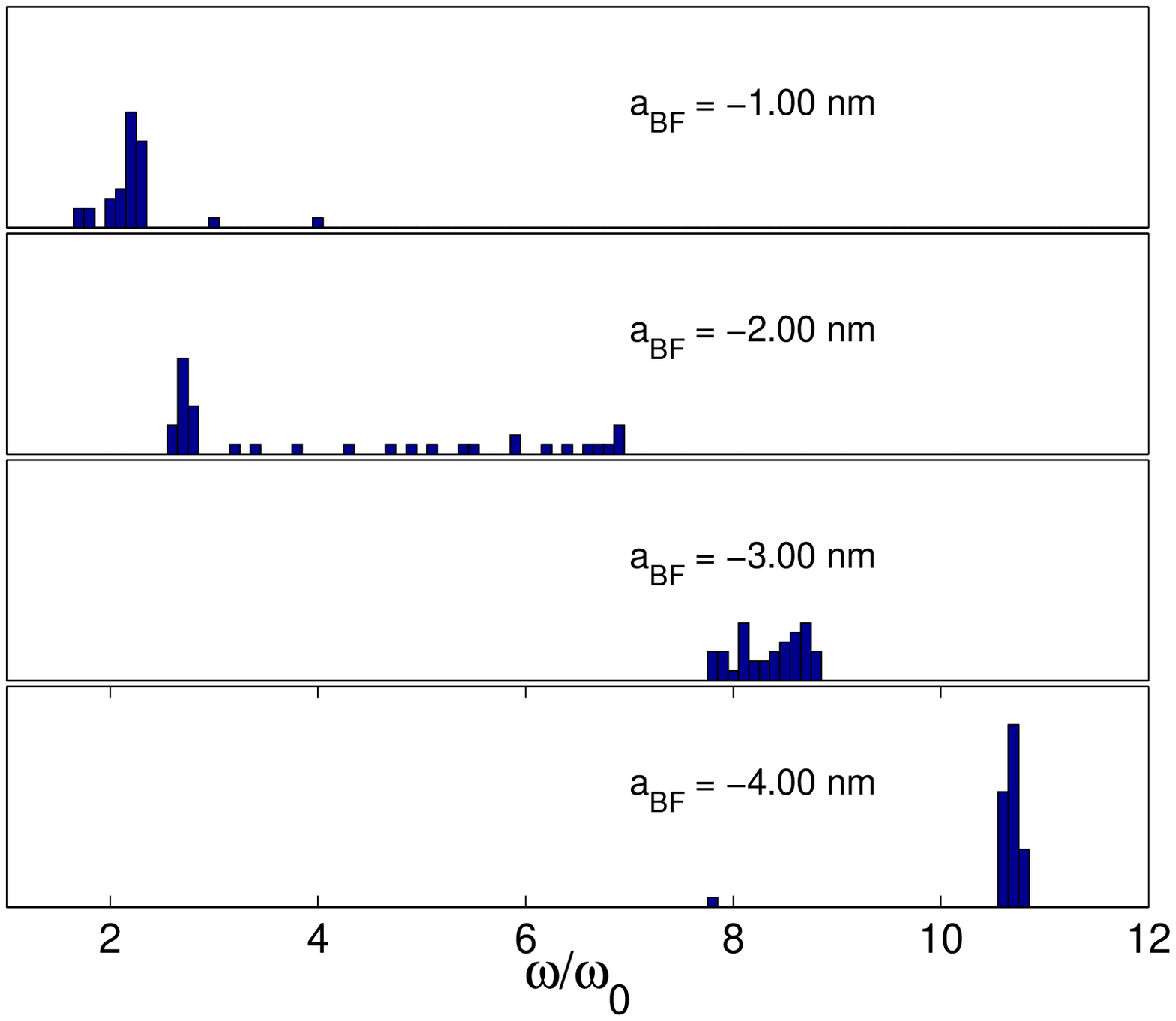}
\includegraphics[width=0.5\linewidth]{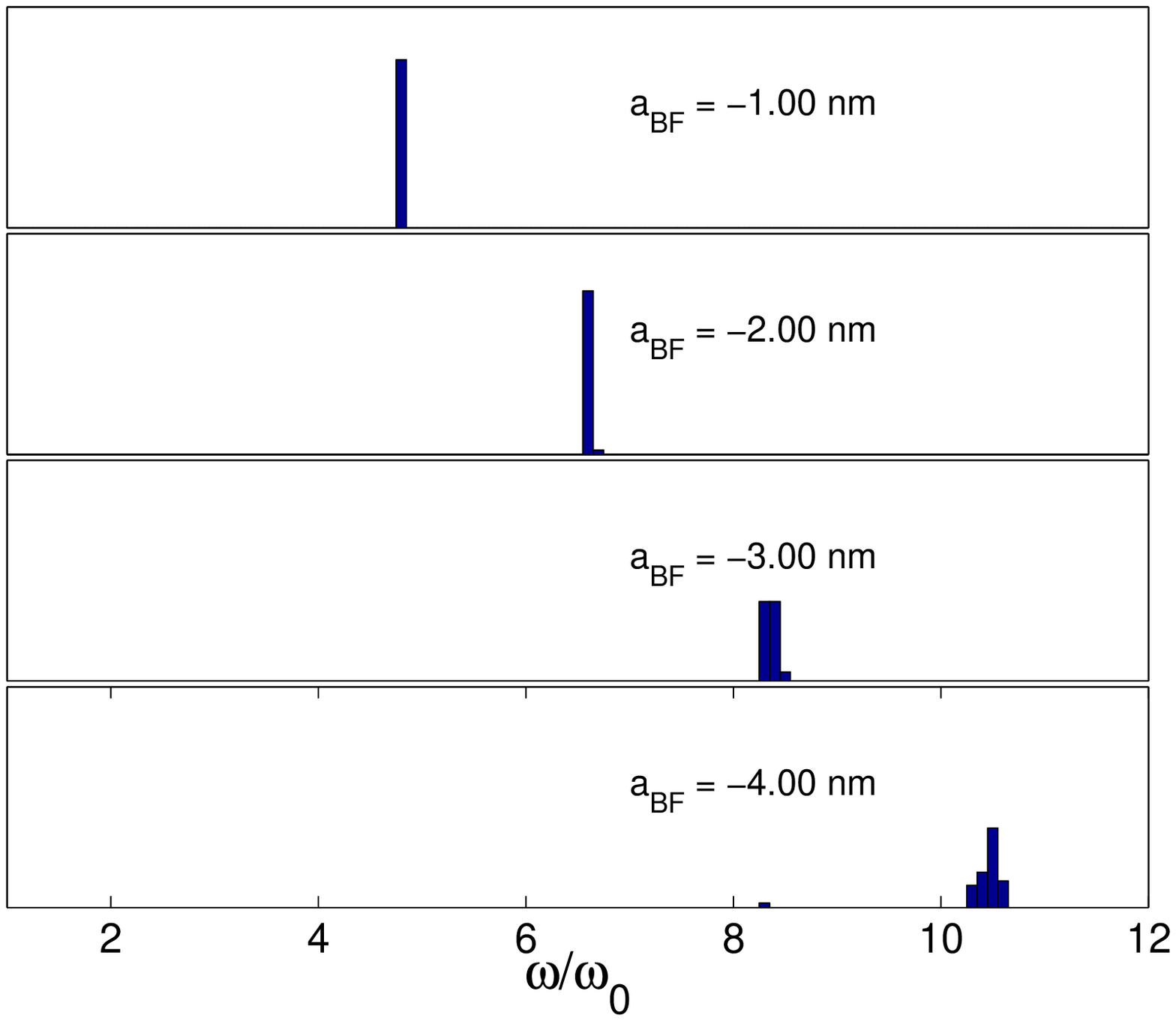}
\caption{\label{fig:hist} Hartree-Fock fermionic spectral functions
 (in arbitrary units) as functions of $\omega/\omega_0$ for $N_F=10^4$
 and $N_B = 10^5$ (top), $10^6$ (middle), and $2.4\times10^7$, at
 various values of $a_{BF}$ as indicated in the panels.}
\end{figure}

\begin{figure}
\includegraphics[width=0.47\linewidth,height=0.5\linewidth]{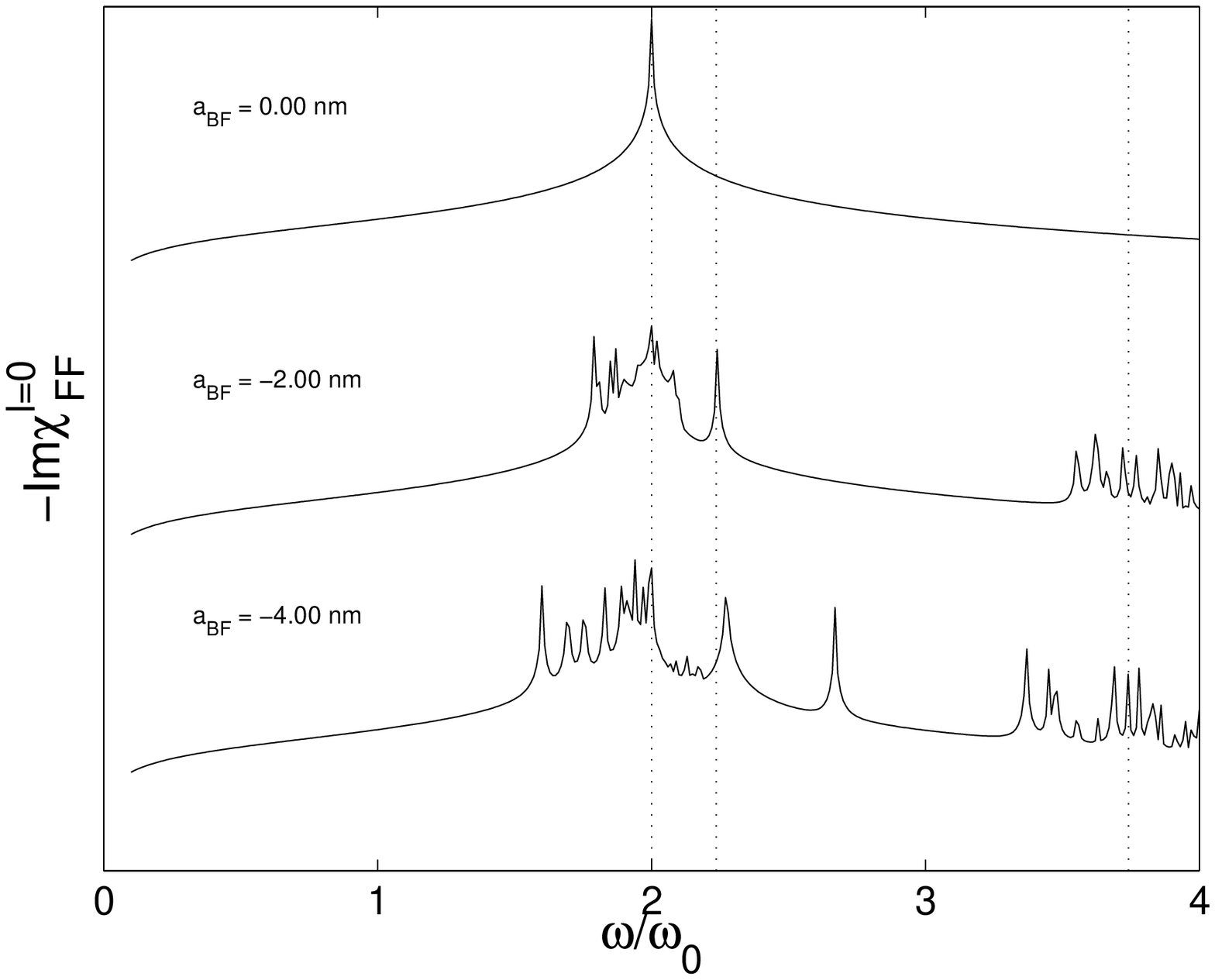}
\includegraphics[width=0.47\linewidth,height=0.5\linewidth]{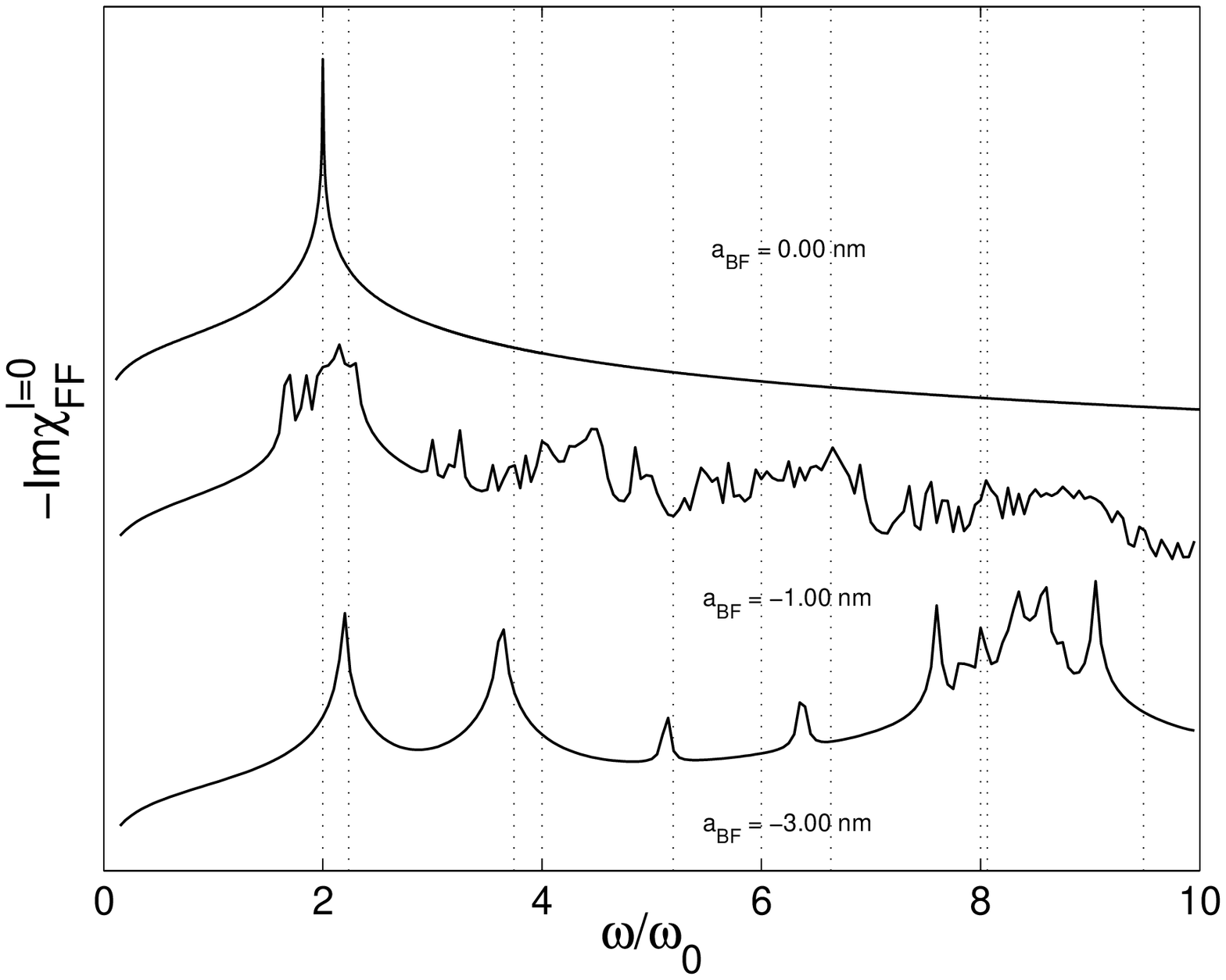}
\caption{\label{fig:spec2} Spectrum of monopolar response (in log
scale and arbitrary units) as a function of frequency $\omega$ (in
units of $\omega_0$) for a boson-fermion mixture in harmonic
confinement with $N_F=10^4$ and $N_B = 10^5$ (left panel) and $10^6$
(right panel), at various values of $a_{BF}$ (in nm) as indicated in
each panel. The vertical dotted lines indicate the frequency of
fermionic (at integer multiples of $2\omega_0$) and bosonic modes in
the absence of boson-fermion interactions.}
\end{figure}

\begin{figure}
\includegraphics[width=0.47\linewidth,height=0.5\linewidth]{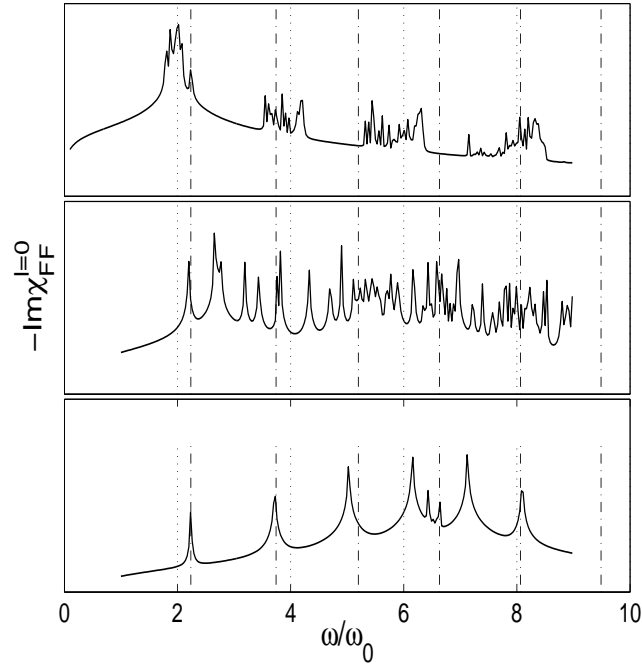}
\caption{\label{fig:spec1} Spectrum of monopolar response (in log
  scale and arbitrary units) as a function of frequency $\omega$ (in
  units of $\omega_0$) for a boson-fermion mixture in harmonic
  confinement with $a_{BF}=-2$ nm and $N_F=10^4$ at various values of
  $N_B$: $N_B=10^5$ (top), $N_B=10^6$ (middle) and $N_B=2.4\times
  10^7$ (bottom).  The vertical lines indicate the frequencies of
  fermionic modes (dotted lines) and bosonic modes (dot-dashed lines)
  in the absence of boson-fermion interactions.  For the above sets of
  parameters thermodynamic collapse is predicted to occur at
  $a_{BF}\simeq - 4.9$ nm for $N_B=10^5$, $-4.4$ nm for $N_B=10^6$,
  and $-4.0$ nm for $N_B=2.4\times 10^7$.}
\end{figure}

An analytical estimate of the compressibility can be obtained by
making a single-pole approximation on the fermion-fermion strength
function
\begin{equation}
{\rm Im} \chi_{FF}(\omega)= -\frac{1}{\pi} A \delta(\omega -\omega_e)
\end{equation}
where for monopolar excitations we take $\omega_e=2 \tilde \omega$.
The oscillator strength $A$ is determined from the $f$-sum rule, which for
a trapped system reads \cite{Bruun01a}
\begin{equation}
-\int \frac{d \hbar\omega}{\pi} \,\hbar \omega\, {\rm
Im}\chi_{FF}(\omega)\equiv M_1=\frac{\hbar^2}{m_F} \int d^3 r |\nabla \delta
U_F(r)|^2 n_F(r) 
\label{fsum}
\end{equation}
In our case $\delta U_F=U_0 (r/a_{ho})^2$ and the RHS of
Eq.~(\ref{fsum}) is determined by the mean square radius of the
equilibrium fermionic density: $M_1=4 \hbar \omega_0 U_0^2 \langle
r^2\rangle/a_{ho}^2 (m_B/m_F)$. This is readily evaluated from the
Thomas-Fermi approximation $n_F(r)=(6\pi^2)^{-1} [(2m_F/\hbar^2)
(\mu_F-m_F\tilde \omega^2r^2/2)]^{3/2}\theta(\mu_F-m_F\tilde
\omega^2r^2/2))$, where for $R_F\simeq R_B$ the chemical potential is
simply given by $\mu_F=\hbar \tilde \omega (6 N_F)^{1/3}$. We
therefore obtain for the oscillator strength the value
\begin{equation} 
A=3^{4/3}2^{-2/3} N_{F}^{4/3}\left(\frac{\omega_0}{\tilde
\omega}\right)^2  U_0^2\left(\frac{m_B}{m_F}\right)^2.
\end{equation}
Finally, from the imaginary part of the strength function we can
calculate the compressibility by means of the Kramers-Kronig
relation, with the result
\begin{equation}
\hbar \omega_0\,  \chi_{FF}(\omega=0)= \frac{3^{4/3}2^{1/3}}{2\pi^2}
N_F^{4/3} U_0^2 \left(\frac{\omega_0}{\tilde
\omega}\right)^3\left(\frac{m_B}{m_F}\right)^2 .
\label{a5}
\end{equation}
This depends on $a_{BF}$ through $\tilde \omega$ and gives a
quantitative agreement with the numerical calculations (see the inset
in Fig~\ref{fig:cmp_static}).
%MASS FACTORS NEED TO BE CHECKED AGAIN U_0-> mF/mB U0%%%%

%\bibliography{pisa,bec}

\newpage

\end{document}